\documentclass[fleqn,usenatbib]{mnras}

\usepackage{newtxtext,newtxmath}

\usepackage[T1]{fontenc}

\DeclareRobustCommand{\VAN}[3]{#2}
\let\VANthebibliography\thebibliography
\def\thebibliography{\DeclareRobustCommand{\VAN}[3]{##3}\VANthebibliography}

\newcommand{\oiii}{[O{\,\scshape iii}]}
\newcommand{\oii}{[O{\,\scshape ii}]}
\newcommand{\oi}{[O{\,\scshape i}]}
\newcommand{\nii}{[N{\,\scshape ii}]}
\newcommand{\sii}{[S{\,\scshape ii}]}

\usepackage[dvipsnames]{xcolor}
\usepackage{tikz,hyperref}

\definecolor{lime}{HTML}{A6CE39}
\DeclareRobustCommand{\orcidicon}{
	\begin{tikzpicture}
	\draw[lime, fill=lime] (0,0) 
	circle [radius=0.13] 
	node[white] {{\fontfamily{qag}\selectfont \tiny ID}};
	\draw[white, fill=white] (-0.0625,0.095) 
	circle [radius=0.007];
	\end{tikzpicture}
	\hspace{-2mm}
}

\foreach \x in {A, ..., Z}{\expandafter\xdef\csname orcid\x\endcsname{\noexpand\href{https://orcid.org/\csname orcidauthor\x\endcsname}
			{\noexpand\orcidicon}}
}

\usepackage{graphicx}	
\usepackage{amsmath}	
\usepackage{appendix}	%
\usepackage{tabularx}
\usepackage{adjustbox}
\usepackage{rotating}
\usepackage{lscape}

\newcommand{\ha}{H$\alpha$ }
\newcommand{\hb}{H$\beta$ }
\newcommand{\has}{H$\alpha$}
\newcommand{\hbs}{H$\beta$}
\newcommand{\ve}{km s$^{-1}$ }
\newcommand{\cbe}{$c$(\hbs) }

\newcommand{\uflux}{erg~cm$^{-2}$~s$^{-1}$}

\graphicspath{
{imagenes/}}

\title[Chemistry and physical properties of HuBi\,1]{Chemistry and physical properties of the born-again planetary nebula HuBi\,1}

\author[Montoro-Molina et al.]{B.\,Montoro-Molina,\thanks{E-mail: borjamm@iaa.es}$^{1\orcidA}$, M.A.\,Guerrero$^{1\orcidB}$, B.\,P\'{e}rez-D\'{i}az$^{1\orcidD}$, J.A.\,Toal\'{a}$^{2\orcidD}$, S.\,Cazzoli$^{1\orcidF}$,
\newauthor M.M.\,Miller Bertolami$^{3,4\orcidE}$ and C.\,Morisset$^{5\orcidG}$
\\
$^{1}$Instituto de Astrof\'{i}sica de Andaluc\'{i}a, IAA-CSIC, Glorieta de la Astronom\'{i}a S/N, Granada 18008, Spain\\
$^{2}$Instituto de Radioastronom\'{i}a y Astrof\'{i}sica, UNAM Campus Morelia, Apartado Postal 3-72, 58090 Morelia, Michoac\'{a}n, Mexico\\
$^{3}$Instituto de Astrof\'{i}sica de La Plata, UNLP-CONICET, La Plata, Argentina\\
$^{4}$Facultad de Ciencias Astron\'{o}micas y Geof\'{i}sicas, UNLP, La Plata, Argentina\\
$^{5}$Instituto de Astronom\'{i}a, Universidad Nacional Aut\'{o}noma de M\'{e}xico, Ensenada, B.C., Mexico
}

\pubyear{2021}

\begin{document}
\label{firstpage}
\pagerange{\pageref{firstpage}--\pageref{lastpage}}
\maketitle

\begin{abstract}
\noindent 
The central star of the planetary nebula (PN) HuBi\,1 has been recently proposed to have experienced a very late thermal pulse (VLTP), but the dilution of the emission of the recent ejecta by that of the surrounding H-rich old outer shell has so far hindered confirming its suspected H-poor nature.  
We present here an analysis of the optical properties of the  ejecta in the innermost regions of HuBi\,1 using MEGARA high-dispersion integral field and OSIRIS intermediate-dispersion long-slit spectroscopic observations obtained with the 10.4m Gran Telescopio de Canarias. 
The unprecedented tomographic capability of MEGARA to resolve structures in velocity space allowed us to disentangle for the first time the H$\alpha$ and H$\beta$ emission of the recent ejecta from that of the outer shell.  
The recent ejecta is found to have much higher extinction than the outer shell, implying the presence of large amounts of dust. 
The spatial distribution of the emission from the ejecta and the locus of key line ratios in diagnostic diagrams probe the shock excitation of the inner ejecta in HuBi\,1, in stark contrast with the photoionization nature of the H-rich outer shell. The abundances of the recent ejecta have been 
computed using the {\sc mappings v} code under a shock scenario.
They are found to be consistent with a born-again ejection scenario experienced by the progenitor star, which is thus firmly confirmed as a new ``born-again'' star. 

\end{abstract}




\section{Introduction}



Planetary nebulae (PNe) are the short-lived descendants of low- and intermediate-mass stars (1~M$_{\sun}$ $\lesssim  M_\mathrm{i} \lesssim$ 8~M$_{\sun}$) in their transition from the asymptotic giant branch (AGB) to the white dwarf (WD) phase. 
At this stage, these stars have completely burnt H and He in their interiors, leaving a $\simeq$0.6 M$_{\sun}$ C-O core. 
After ejecting most of their H-rich envelopes, the now post-AGB star evolves increasing its surface temperature at roughly constant luminosity burning the rest of its envelope until H is exhausted. 
The central star (CSPN) enters then the WD cooling track, irradiating thermal energy and cooling down.

A selected group of PNe can experience a ``second life''. During the post-AGB evolution, their CSPNe can build up a He mantle, reaching the conditions for a very late thermal pulse \citep[VLTP,][]{Schonberner1979,Iben}, which converts He into C and O and ejects highly-processed H-deficient material at high-speeds into the old H-rich PN \citep[][]{Miller}. As the stellar envelope expands, it cools dramatically, pushing the CSPN back to lower effective temperatures ($T_\mathrm{eff}$), first as a H-deficient giant, now as a late C-rich Wolf-Rayet ([WC]) type star.  
In a sense, the CSPN is born-again. 
The new PN emerging inside the old PN will be also referred here as a born-again PN.

HuBi\,1 (a.k.a.\ PM\,1-188) has been one of the latest PN to be proposed to belong to the class of born-again PNe \citep[][]{Guerrero2018}. 
This object has been gaining attention over the years, demonstrating to have a unique late evolutionary behaviour.
\citet{HuBibo} noticed it to be a low-excitation PN whose CSPN, the bright IR source IRAS\,17514$-$1555, has a late [WC] spectral type.
According to \citet{Pollacco}, this PN consists of a faint extended low-density outer shell with typical PN abundances and an unresolved inner shell with apparent bipolar structure\footnote{
\citet{Pollacco} also reported an extremely high density for this inner shell, but it resulted from the contamination of the density sensitive [S~{\sc ii}] $\lambda\lambda$6716,6731 doublet by stellar C~{\sc ii} emission lines. 
}. 
Discrepancies between the observed electron density ($n_\mathrm{e}$ $\lesssim$1000~cm$^{-3}$) and that much higher expected for a nebula around a late [WC] CSPN led \citet{Pena2001} to propose either a born-again scenario for HuBi\,1 or the slow evolution of a low-mass AGB star.  
The chemical abundances derived at that time \citep{Pollacco,Pena2005} were somehow similar to that of typical Galactic PNe, although the bright He~{\sc i} $\lambda$5876 emission line suggested a large He/H abundance ratio. 
On the other hand, the low effective temperature $T_\mathrm{eff} \simeq$35,000~K of the CSPN \citep{LH1998} rather supported the latter scenario \citep{Pena2005}.  



It has not been until recently that the born-again scenario in HuBi\,1 has gained momentum. 
\citet{Guerrero2018} noticed a decrease in the CSPN brightness $\simeq$10 mag in the last 46 yr that was attributed to a circumstellar veil of C-rich dust suggested by the presence of numerous carbon lines in the optical spectrum of the CSPN \citep{Pollacco} and by the strong IR emission \citep{HuBibo}. 
Such C-rich material would have been ejected through a VLTP event and, as it expands and cools down, reaches optimal conditions for its condensation on dust grains \citep[e.g.,][]{Perea2009}.

\citet{Guerrero2018} proposed that the decreasing ionizing flux of the CSPN causes the recombination of the outer shell, whereas the 
detection of He~{\sc ii} emission from the inner shell and its atypical inverted ionisation structure, with He~{\sc ii} emission embracing the emission of [O~{\sc iii}] and [N~{\sc ii}], rather suggest it is shock-heated. 
Indeed, the detection of material expanding up to $\simeq$300 km~s$^{-1}$ in the innermost region of HuBi\,1 \citep{Rechy} lend strong support to the shock excitation of this region by a 
recent ejecta.  



The most convincing evidence to fully declare HuBi\,1 to be part of the class of born-again PNe would come from the chemical abundances of the recent ejecta, since noticeable abundances discrepancies are expected in a born-again event. H-poor material and enhanced C, N and O abundances would be present in the most recent ejecta produced by the VLTP \citep{Miller} with the outer nebula exhibiting ``normal'' PNe abundances. 
However, \citet{Pena2021} have recently reported no abundances differences between the inner and outer shells of HuBi\,1, in accordance with previous studies. 
These results ought to be questioned as they are based on the assumption of the photoionisation of the inner shell, which does not seem to be the case given the evidence for shock excitation.  
Most importantly, since these authors did not remove the contamination of the bright H Balmer emission lines from the outer shell to the inner shell, their analysis of its spectrum can be expected to artificially enhance its H content.



In this paper we combine optical high-dispersion integral field spectroscopy (IFS) and intermediate-dispersion long-slit spectroscopic observations of HuBi\,1 to isolate the emissions of its inner and outer shells in order to determine the H, He, N, and O abundances of the inner shell.  
The observations and data reduction are presented in Section~2.  
The data analysis is discussed in Section~3, where two appendices are presented in order 
(i) to assess the contribution of the emission from the outer shell to the emission of the inner shell, and 
(ii) to correct the effects of the atmospheric differential chromatic refraction that affects the long-slit spectroscopic observations\footnote{
\citet{Pena2021} used the same long-slit spectroscopic data set presented here, but did not apply any correction for atmospheric differential chromatic refraction.  
The correction of these effects will be shown to be essential for an appropriate determination of the physical properties and chemical abundances of the inner shell of HuBi\,1.
}. 
The results are presented in Section~4, including maps of extinction and electron density of the inner shell, the assessment of the importance of shocks and ultimately the calculation of the chemical abundances. 
Finally, the discussion and summary are provided in Sections~5 and 6, respectively.  

\section{Observations and data reduction}

\subsection{GTC MEGARA IFS observations}

IFS observations of HuBi\,1 were obtained on 2020 August 6 (Program ID GTC96-20A, PI Guerrero) using the Multi-Espectrógrafo en GTC de Alta resolución para Astronomía \citep[MEGARA;][]{GildePaz2018} at the 10.4 m Gran Telescopio Canarias (GTC). 
The high-resolution (HR) Volume-Phased Holographic (VPH) grism VPH665 and the medium-resolution (MR) grism VPH481 were used. 
\begin{figure}
\centering
\includegraphics[clip,width=1\columnwidth]{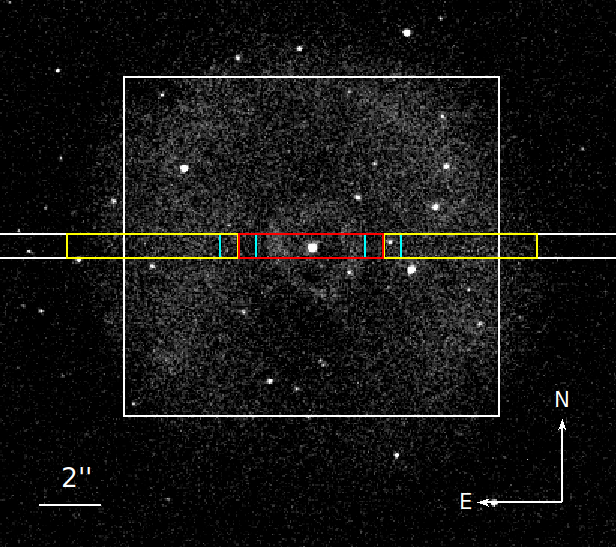}
\caption{
\emph{HST} WFPC2 F656N image of HuBi\,1. 
The positions of the 12\farcs5$\times$11\farcs3 FoV of MEGARA's IFU and the 0\farcs8-width OSIRIS long-slit are overlaid using white lines lines. 
This \emph{HST} image reveals clearly the position of background stars included in the apertures of these instruments. 
The apertures used for extraction of 1D spectra in the OSIRIS data for the inner and outer regions, and from an additional intermediate region are overlaid in red, yellow, and cyan lines, respectively. 
}
\label{fig:fov_int}
\end{figure} 

\begin{figure*}
\centering
\includegraphics[width=0.25\linewidth]{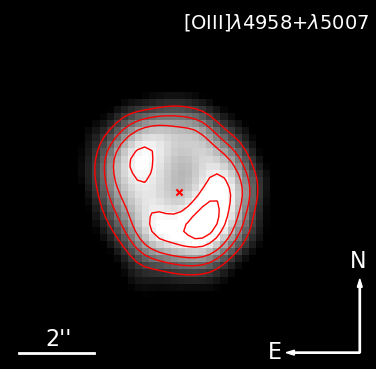}~
\includegraphics[width=0.25\linewidth]{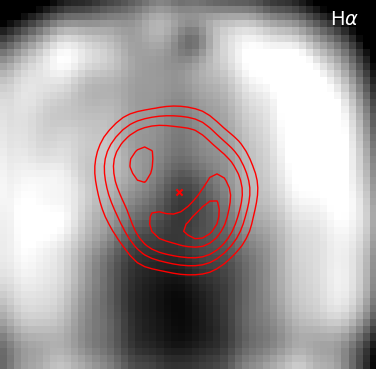}~
\includegraphics[width=0.25\linewidth]{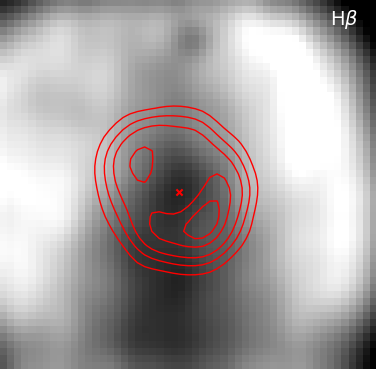}~
\includegraphics[width=0.25\linewidth]{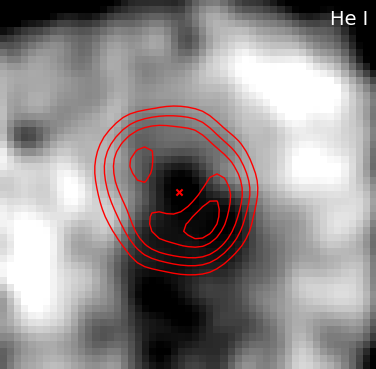}\\
\includegraphics[width=0.25\linewidth]{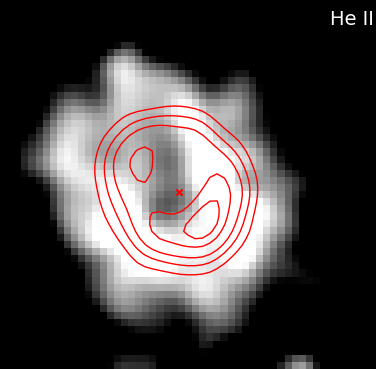}~
\includegraphics[width=0.25\linewidth]{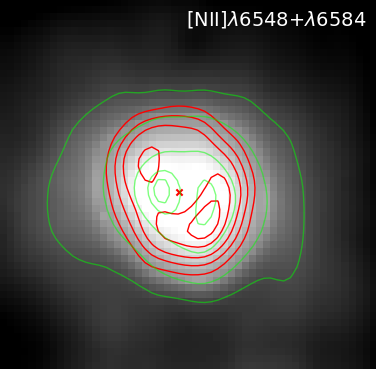}~
\includegraphics[width=0.25\linewidth]{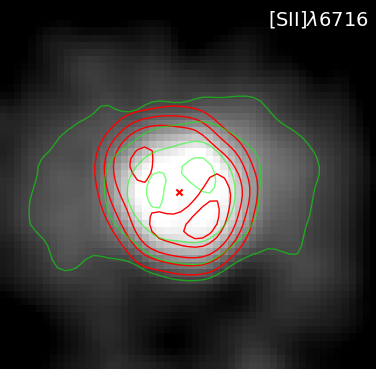}~
\includegraphics[width=0.25\linewidth]{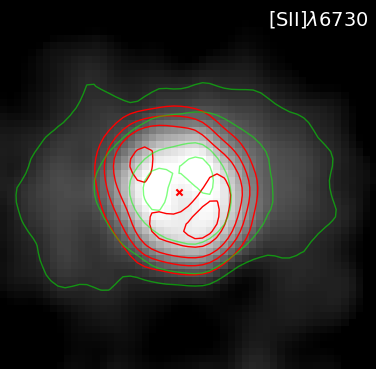}
\caption{
GTC MEGARA maps of HuBi\,1 in different emission lines. 
From left to right, the panels represent the maps of 
[O~{\sc iii}] $\lambda\lambda$4959,5007 \r{A}, H$\alpha$, H$\beta$, and He~{\sc i} $\lambda6678$ \r{A} (top), and 
He~{\sc ii} $\lambda4686$ \r{A}, [N~{\sc ii}] $\lambda\lambda$6548,6584 \r{A},  
[S~{\sc ii}] $\lambda6716$ \r{A}, and [S~{\sc ii}] $\lambda6730$ \r{A} (bottom). 
To facilitate the comparison among the different images, red contours extracted from the [O~{\sc iii}] image tracing the inner shell are overlaid on all images. 
Green contours are also overlaid in the [N~{\sc ii}] and [S~{\sc ii}] images to emphasise the emission from these lines from an intermediate region external to the inner shell. 
The red contours have been selected at 
66\%, 55\%, 44\%, 33\%, and 22\% of the [O~{\sc iii}] brightness peak, whereas the green contours have been selected at 
75\%, 70\%, 55\%, 25\%, and 15\% of the [N~{\sc ii}] brightness peak, and 
95\%, 80\%, 55\%, and 40\%, of the [S~{\sc ii}] brightness peak.
}
\label{fig:total_flux}
\end{figure*}

The observations consisted of three 900~s HR-VPH and six 1000~s MR-VPH exposures obtained under optimal transparency conditions and excellent seeing 0\farcs8 as measured at the DIMM station. 
The VPH665 grism provides a spectral dispersion of 0.093 \r{A}~pix$^{-1}$ and a full-width at half-maximum (FWHM) spectral resolution $\approx$ 15 km s$^{-1}$, that is, $R \approx$ 20050, while the VPH481 grism provides a spectral dispersion of 0.105 \r{A}~pix$^{-1}$ and FWHM spectral resolution $\approx$ 23 km s$^{-1}$, i. e., $R \approx 13200$. 
The spectral ranges 4585.7--5025.1 \r{A} and 6405.6--6797.1 \r{A} covered by these two VPHs contain the key emission lines of He~{\sc ii} $\lambda4686$ \r{A}, H$\beta$, [O~{\sc iii}] $\lambda\lambda4959,5007$ \r{A}, [N~{\sc ii}] $\lambda\lambda6548,6584$ \r{A}, H$\alpha$, He~{\sc i} $\lambda6678$ \r{A}, and [S~{\sc ii}] $\lambda\lambda6716,6730$ \r{A}. 
The integral-field unit (IFU) mode was used. 
It has 567 hexagonal spaxels with a diameter of 0\farcs62 and a field of view (FoV) of 12\farcs5$\times$11\farcs3, as shown in Figure~\ref{fig:fov_int}.

The raw MEGARA data were reduced following the Data Reduction Cookbook \citep{Pascual2019} using the \textit{megaradrp} v0.10.1 pipeline released on 2019 June 29. 
This pipeline applies sky and bias subtraction, flat-field correction using halogen internal lamps, 
wavelength calibration, and spectra tracing and extraction. 
We note that there is an appreciable illumination pattern making brighter the East and West edges of the FoV, which reduced the cube useful FoV to a region 11\farcs3$\times$11\farcs3 in size. 
The result is a row-stacked spectrum, which is converted into a 52$\times$58 map of 0\farcs2 square spaxel on spatial dimension and 4300 spaxel on spectral axis using the regularization grid task \textit{megararss2cube}. 
The flux calibrations were performed using observations of the spectro-photometric standard HR\,7596 obtained immediately after those of HuBi\,1.

\subsection{GTC OSIRIS long-slit spectroscopic observations}

Intermediate resolution long-slit spectroscopy was acquired on 2018 May 14 (Program ID GTC112-18A, PI Guerrero) with the Optical System for Imaging and low-intermediate-Resolution Integrated Spectroscopy (OSIRIS)\footnote{\url{http://www.gtc.iac.es/instruments/osiris/media/OSIRIS-USER-MANUAL_v3_1.pdf}} at the GTC. 
The observation consisted of 4 exposure of 1575 s on long-slit spectroscopy mode. 
The R1000B grism was used, providing a spectral dispersion of 2.12 \r{A} pix$^{-1}$ in the spectral range 3630--7500 \r{A} at a spectral resolution  $R \approx 1120$.
The slit position was placed at a position angle (PA) 90$^\circ$ across the central star (see Fig.~\ref{fig:fov_int}).

The raw data were bias and sky subtracted and flat-field corrected following standard routines in {\sc iraf} \citep{Tody1993}. 
The wavelength calibration was carried out using Ne and HgAr lamps, whereas the flux calibration was obtained using observations of the spectro-photometric standard star GD\,140 obtained in the same night.

The relative flux calibration of the GTC MEGARA and OSIRIS data was checked by comparing the flux of a number of bright emission lines measured from identical spatial apertures in both data sets. 
These fluxes were found to agree within a few percent.

\section{Data Analysis}

The OSIRIS long-slit spectra and MEGARA IFS observations provide a wealth of information on the spatially-resolved spectral properties of HuBi\,1 that has allowed us to investigate the ionisation structure and extract clean spectra of its inner shell.  
These are described in the next sections.

\subsection{The inverted ionisation structure of HuBi\,1}

The final MEGARA data cube allows the extraction of background-subtracted narrow-band images in emission lines of interest, thus providing great versatility to investigate the complex structure of HuBi\,1. 
The data analysis method is as simple as selecting appropriate ranges of spectral channels encompassing an emission line to build a background-subtracted narrow-band image of that emission line. \\
The images of the total integrated flux of key emission lines extracted from the MEGARA data presented in Figure~\ref{fig:total_flux} confirm the remarkable inverted ionisation structure of HuBi\,1.

The leftmost-top panel of Figure~\ref{fig:total_flux} presents the [O~{\sc iii}] emission, which was obtained by adding the [O~{\sc iii}]~4959~\AA\, and [O~{\sc iii}]~5007~\AA\, emission lines to improve the signal-to-noise ratio (SNR). 
This emission has a ring-like appearance with an averaged extension of $\sim$2\farcs5 in radius.  
The emission is elongated along the NE direction, with brighter emission at the tips, as shown by the [O\,{\sc iii}] contours overplotted. 
The [O\,{\sc iii}] emission lines uniquely trace the inner region of HuBi\,1 and thus [O\,{\sc iii}] contours will be overplotted on the other key emission line maps of Figure~\ref{fig:total_flux} for comparison.

The images in the H$\alpha$, H$\beta$ and He~{\sc i}~6678~\AA\, emission lines in Figure~\ref{fig:total_flux}, on the other hand, trace the outer nebular shell of HuBi\,1. 
Unfortunately, the MEGARA FoV does not cover the full extent of this shell, which has an averaged diameter of $\sim$15~arcsec. 
The inner shell of HuBi\,1 shown by the red [O~{\sc iii}] contours is encompassed by the emission in the H$\alpha$, H$\beta$ and He\,{\sc i} lines.  
It is not possible, however, to assess whether the inner shell emits in these emission lines or whether the emission detected inside the [O\,{\sc iii}] contours of the inner shell in these images is rather due to projection effects.

Maps of the low-ionisation [N~{\sc ii}] and [S~{\sc ii}] emission lines are also presented in Figure~\ref{fig:total_flux}, where green contours have been used to highlight the distribution and extent of their emission.  
The bulk of the emission in these low-ionisation lines is confined within the red [O~{\sc iii}] contours that define the inner shell, with two peaks separated by $\sim$1\farcs5 aligned mostly along the East-West direction, i.e. they are clearly misaligned with the peaks in the [O~{\sc iii}] image.  
These images also reveal the presence of an intermediate region\footnote{
Here the emission from the [N~{\sc ii}] $\lambda$6548 \AA\ and [N~{\sc ii}] $\lambda$6584 \AA\ emission lines has been added to enhance the weak emission from the intermediate region, whereas the [S~{\sc ii}] lines are presented separately because the smaller intensity contrast of the emission from the inner shell and intermediate region allows an easier view of the latter.
} 
spreading out $\sim$4~arcsec the emission of the inner shell into the outer shell along the East-West direction.


Finally, the leftmost-bottom panel of Figure~\ref{fig:total_flux} shows the spatial distribution of the emission in the line of He~{\sc ii} $\lambda$4686 \AA, the one with the highest ionisation potential in the nebular spectrum of HuBi\,1. 
The emission from the inner shell in this emission line is the most extended, with an angular radius of $\sim$4$\times$3~arcsec oriented along a PA similar to that of the emission in the [O~{\sc iii}] lines. 
The He~{\sc ii} emission encircles that of the [O~{\sc iii}], which 
surrounds that of the [N~{\sc ii}] and [S~{\sc ii}] emission lines. 
The inverted ionisation structure of the inner shell of HuBi\,1, suggested originally by \citet{Guerrero2018} using 1D line emission spatial profiles, is therefore firmly confirmed in the MEGARA 2D emission maps shown in Figure~\ref{fig:total_flux}.


\subsection{Dissecting the inner shell of HuBi\,1 with MEGARA}
\label{sec:sep_shell}

The emission of the inner shell of HuBi\,1 in the images presented in Figure~\ref{fig:total_flux} would generally include a contribution from the outer shell, as the former is seen through the latter. 
The emission from the different shells of HuBi\,1 can be kinematically separated given their distinct expansion velocities \citep{Guerrero2018,Rechy,Pena2021} taking advantage of the MEGARA IFS high spectral resolution, although the quality (and reliability) of the kinematic separation of the emission from the two shells in a given emission line would depend certainly on the spatio-kinematic distribution of that line.  
A detailed description of the different types of lines according to the varying relative contributions of the different structures of HuBi\,1 in each emission line is given in Appendix~\ref{sec:OSIRIS_data}.  
These are briefly discussed in the following.

First, the images of the [O~{\sc iii}] and He~{\sc ii} emission lines can be attributed purely to emission from the inner nebula. 
We will refer to them as purely internal (PI; see Appendix~\ref{sec:OSIRIS_data}) emission lines. 
On the other hand, the emission of He~{\sc i} arises uniquely from the outer shell and this will accordingly be referred as a purely external (PE; see Appendix~\ref{sec:OSIRIS_data}) emission line.

There are then a number of emission lines whose images definitely include emission from both the inner and outer shells.  
The images of the emission from the [N~{\sc ii}] and [S~{\sc ii}] lines is mostly dominated by the emission from the inner shell, but there is a non-negligible contribution of emission arising from regions outside the inner shell. 
These lines are thus referred as internal lines with emission from the external shell (IwE; see Appendix~\ref{sec:OSIRIS_data}). 
We note that the spatial distribution of the outer emission in these lines do not follow the shell-like morphology of the emission in the \hbs, \ha and He~{\sc i} emission lines, but it is rather spread between the inner and outer shells in a sort of intermediate region. 
The spatio-kinematic distribution of the emission in these lines from this intermediate region does not allow a clear separation from the emission of the inner shell \citep[see discussion section in][]{Jesus}.

Finally, the emission of the \hb and \ha Balmer lines in the spectral range covered by the MEGARA observations are dominated by the outer shell emission, but the emission from the inner shell can be still kinematically resolved. 
These lines are referred in Appendix~\ref{sec:OSIRIS_data} as external lines with inner emission (EwI). 
Assessing the specific H$\alpha$ and H$\beta$ fluxes of the inner and external shells will help us accurately unveil the extinction contribution and abundances for each shell. 
The emission from the inner shell in these two emission lines will be extracted applying two different methods in the next two subsections.  

\begin{figure}
\includegraphics[width=0.5\linewidth]{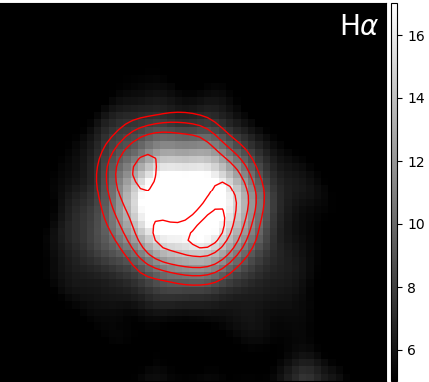}~
\includegraphics[width=0.5\linewidth]{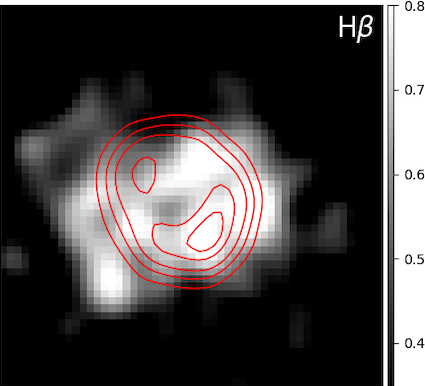}\\
\includegraphics[width=0.5\linewidth]{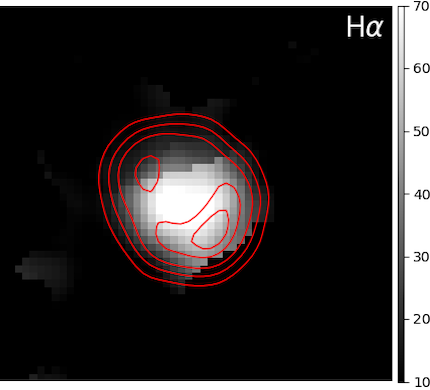}~
\includegraphics[width=0.5\linewidth]{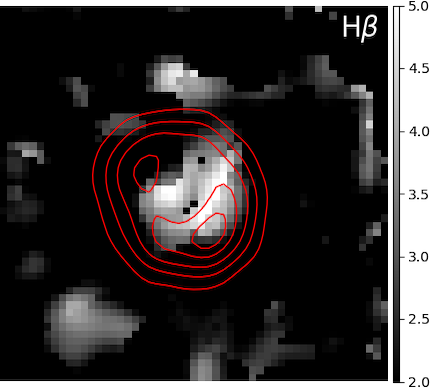}
\caption{
GTC MEGARA H$\alpha$ and H$\beta$ images of the inner region of HuBi\,1 corresponding to the high-velocity components in these emission lines as derived from the isolation of the MEGARA channels within a velocity interval (top panels, see Sec.~\ref{sec:isolation}) and from the best multi-Gaussian fits (bottom panels, see Sec.~\ref{sec:isolation-gaussian}).
The extent of the inner shell is shown for comparison using the red contours derived from the [O\,{\sc iii}] image in Figure~\ref{fig:total_flux}. 
The side scale shows the surface brightness in units of $10^{-17}$ erg~cm$^{-2}$~s$^{-1}$~arcsec$^{-2}$.
}
\label{fig:maps_int}
\end{figure}

\subsubsection{High-velocity component isolation in the \hb\ and \ha\ emission}
\label{sec:isolation}

The inner shell of HuBi\,1 can be described as a fast expanding shell-like structure or outflow with expansion velocities up to $\simeq$300~km~s$^{-1}$, well exceeding that of the outer shell $\lesssim$50 \ve \citep{Rechy}. 
The high-velocity emission from the inner shell is indeed revealed as red and blue wings in the \hb and \ha line profiles in the velocity ranges from $-14$ to $+20$ \ve and $+108$ to $+122$ \ve for \hbs, and from $-96$ to $+11$ \ve and $+109$ to $+171$ \ve for \has.  
The differences in the velocity ranges of the wings of \ha and \hb can be attributed to the larger SNR of the former emission line. 


The spectral channels of the MEGARA data cube corresponding to the velocity intervals described above have been added to obtain the images of the inner shell in these emission lines shown in the top panels of Figure~\ref{fig:maps_int}.  
The \ha map (top-left panel of Fig.~\ref{fig:maps_int}) is notably less noisy than the \hb one (top-right panel of Fig.~\ref{fig:maps_int}) due to the better spectral resolution of the HR-VPH and higher SNR of the \ha line.  
The emission in this \ha map is closely inscribed within the emission of the inner shell represented by the red contours extracted from the [O~{\sc iii}] image (leftmost-top panel of Fig.~\ref{fig:total_flux}). 
Both the \ha\ and \hb\ maps peak at the Southwest tip of the inner shell, as it does the [O~{\sc iii}] emission from the inner shell.  
Total integrated fluxes of $8.1\times10^{-15}$ \uflux\ in \ha\ and $4.8\times10^{-16}$ \uflux\ in \hb\ were derived adding the flux in all pixels within a circular area 2\farcs5 in radius.



\subsubsection{Multi-Gaussian fit to the H$\alpha$ and H$\beta$ emission lines}
\label{sec:isolation-gaussian}

\begin{figure}
\includegraphics[width=\linewidth]{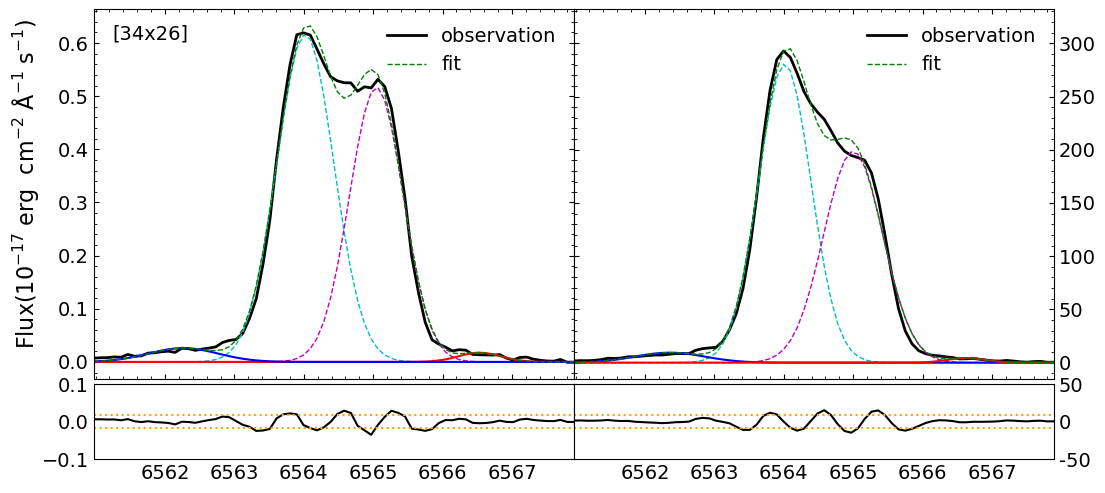}\\
\includegraphics[width=\linewidth]{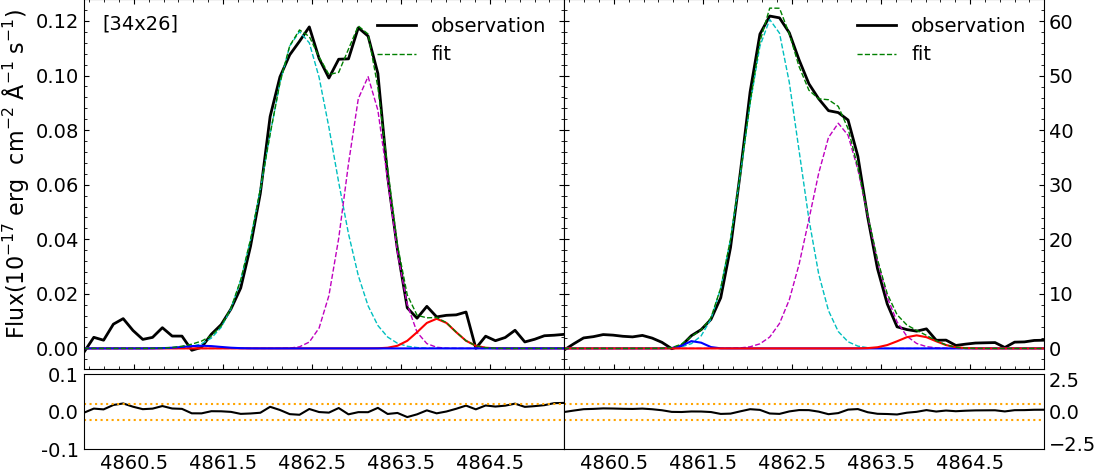}
\caption{
Examples of \ha (top) and \hb (bottom) multi-Gaussian fits 
(\emph{left}) for the spectrum of the pixel (34,26), which is offset 1\farcs3 along PA 285$^\circ$ from the centre at pixel (28,25), and (\emph{right}) for the integrated spectrum of the inner shell, extracted  from a circular aperture 2\farcs6 in radius.
The solid black line corresponds to the observed line profile, whereas the green dotted lines represents the multi-Gaussian fit to this profile.  
The blue, cyan, magenta, and red solid lines correspond to the individual Gaussian components for the fast approaching component of the inner shell, the approaching and receding sides of the outer shell, and the fast receding component of the inner shell, respectively. 
The bottom panels show the residuals of the fit in solid black lines, whereas the dotted horizontal orange lines correspond to 3$\sigma$ of the continuum noise derived from representative spectral ranges at both sides of the \ha and \hb lines, respectively.
}
\label{fig:fit_example}
\end{figure}

Although the method described in Sec.~3.2.1 has proven effective to isolate the emissions of the \ha and \hb lines from the inner shell and to estimate their fluxes, it has some shortcomings. 
In particular, the selection of the spectral channels (or velocity intervals) mapping the emission of the inner shell is affected by the differing spectral resolution and SNR of the \hb and \ha lines. 
As a result, the \hb map is derived from a velocity interval shorter than that of the \ha map, which may have effects on the \has/\hb ratio and on the estimate of the logarithmic extinction coefficient c(\hbs) of the inner shell based on that ratio.  
Moreover, the channels corresponding to the low-velocity tail of the emission from the inner shell had to be discarded because they were actually dominated by emission from the much brighter outer shell, which may lead to an underestimation of the flux from the inner shell. 
We remind the reader that one of the main objectives of this work is the determination of the \ha (or \hbs) flux to assess whether the chemical abundances of the inner shell of HuBi\,1 are consistent with those of a born-again nebula. 
It is thus critical to obtain reliable values of the \hb and \ha fluxes for the inner shell.

Therefore we have explored the determination of these fluxes using multi-Gaussian fits at each pixel onto the inner shell of HuBi\,1 to excise the emission of the dominant outer shell from that of the inner shell \citep[see a similar approach in the case of M\,2-31,][]{Rechy2021}. 
As described above, this is possible because the inner shell of HuBi\,1 has a much larger expansion velocity than the outer shell \citep{Rechy}.  
The distinct kinematic behavior of the inner and outer shells reflects in the spectral profiles of the \ha and \hb emission lines at the location of the inner shell (approximately the innermost 2\farcs5 in radius central region), where up to four velocity components can be found; two from the main nebula with slow velocity (one approaching and one moving away) and another two from the inner shell that are detected as high-velocity red and blue wings.

An example of multi-Gaussian fits to the \ha and \hb emission lines for a pair of pixels projected onto the inner shell\footnote{
The outer shell was also fitted with one, two or up to three components, depending on the spatial location, for continuity with the fit to the inner shell. 
} 
is shown in Figure~\ref{fig:fit_example}. 
These are overplotted with the best multi-Gaussian fits revealing the presence of weak red and blue wings corresponding to the inner shell.  
The small flux contribution from these components might have escaped the previous determination of the \ha and \hb fluxes. 
After applying this method to all pixels, the resulting total flux is $1.6 \times 10^{-14}$ \uflux\ in \ha and $1.1 \times 10^{-15}$ \uflux\ in H$\beta$, i.e., about two times the flux derived by the previous method. 
Maps with the the best multi-Gaussian fits for each pixel are presented in the bottom panels of Figure~\ref{fig:maps_int} for comparison with those obtained in the previous subsection.


\subsection{OSIRIS spectra of HuBi\,1}

The GTC OSIRIS long-slit spectrum, with a broader spectral coverage than the MEGARA data, has been used to extract one-dimensional spectra of the inner and outer shells of HuBi\,1. The emission of the inner shell was extracted from a 5$^{\prime\prime}$ in length central aperture of the OSIRIS two-dimensional spectrum covering the extent of the [O~{\sc iii}] $\lambda5007$ emission in the MEGARA maps (the red rectangular aperture in Fig.~\ref{fig:fov_int}).  
The emission of the outer region was extracted from two apertures of $\sim$5$^{\prime\prime}$ in length at both sides of the central region (the yellow rectangular apertures in Fig.~\ref{fig:fov_int}) and then added together.  
The emission of a transition region between the inner and outer shells was also extracted from two small apertures 1$^{\prime\prime}$ in length (the cyan rectangular apertures in Fig.~\ref{fig:fov_int}) to assess the variation of key emission lines from one region to the other.

\begin{figure*}
\includegraphics[width=1\linewidth,height=0.5\textheight]{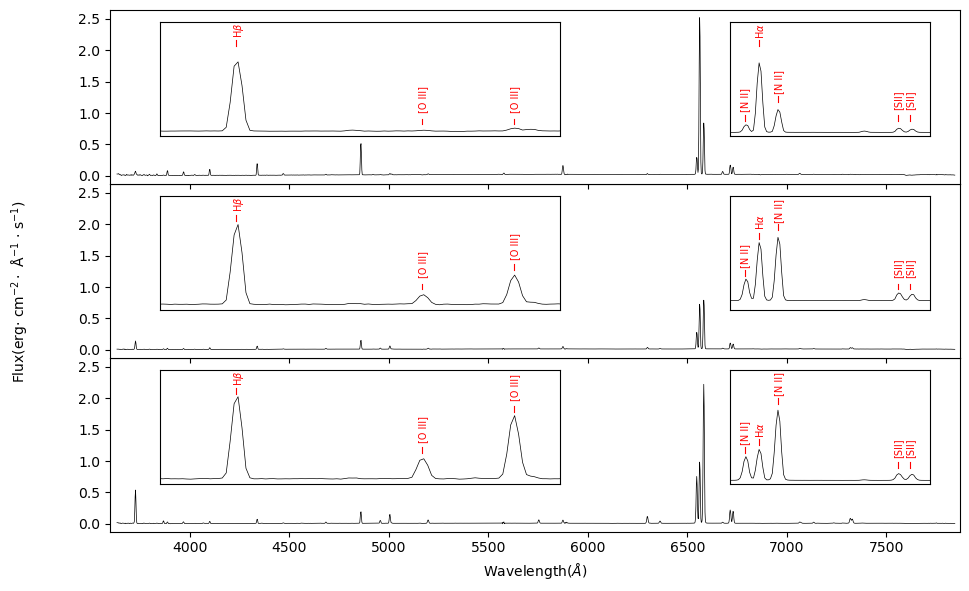}
\caption{
OSIRIS intermediate-dispersion spectra of the outer shell (top), intermediate region (middle), and inner shell (bottom) of HuBi\,1 extracted from the yellow, cyan, and red apertures shown in Figure~\ref{fig:fov_int}, respectively. 
Two spectral regions of interest are zoomed to reveal the notable variations in the key emission lines among the different shells.
}
\label{fig:spec_osi}
\end{figure*}

The resulting spectra are shown in Figure~\ref{fig:spec_osi}, with insets to zoom in two spectral regions of interest. 
These spectra reveal notable spectral variations among the different regions. 
The [O~{\sc iii}] $\lambda$5007 emission line is almost as intense as the \hb line in the inner region (bottom panel of Fig.~\ref{fig:spec_osi}) but, as we moved to more external regions, it fades and mostly disappears (top panel in Fig.~\ref{fig:spec_osi}). 
On the other hand, the [N~{\sc ii}] $\lambda$6584 emission line prevails over the \ha line in the inner shell, while in the outer shell their relative importance is reversed \citep[see figure~8 in][for a similar case for the born-again PN A\,58]{Guerrero1996}. 
These variations in line ratios imply an anomalous ionisation structure. 

A standard analysis of these spectra involves the measurement of the relative line intensities \citep[e.g.,][]{Pena2021}, but it must be noted that the GTC OSIRIS spectrum of HuBi\,1 was acquired at a low elevation, when the effects of differential chromatic refraction (DCR) along the parallactic angle cannot be neglected for data obtained over such a broad spectral range \citep{Filippenko1982}.  
This is a problem particularly serious for these observations, whose slit is almost orthogonal to the parallactic angle. 
As a result, the apparent position of the slit on the sky for emission lines at different wavelengths will shift noticeably.  
The details and extent of these effects and a solution using the OSIRIS data in conjunction with the MEGARA data to measure the relative line intensities of the emission lines in the OSIRIS spectra are described in detail in Appendix~\ref{sec:chromatic}.

\begin{table}
\caption{
Relative fluxes and line intensities with respect to H$\beta$ (=100) of the emission lines of HuBi\,1 measured from OSIRIS and MEGARA data. 
Emission lines from OSIRIS for which MEGARA data are available have been replaced by the later. 
Due to the DCR effects affecting the OSIRIS data, each line of this set is normalized by the \hb flux computed for the equivalent aperture derived from the MEGARA data (see Appendix A). 
}
\setlength{\tabcolsep}{0.65\tabcolsep}
\label{tbl:t_f_OSI}
\begin{tabular}{lrrrr}
\hline
Line & 
\multicolumn{2}{c}{\underline{~~~~~~~~Inner shell~~~~~~~~}} & 
\multicolumn{2}{c}{\underline{~~~~~~~~Outer shell~~~~~~~~}} 
\\
 & 
\multicolumn{1}{c}{$F$} & 
\multicolumn{1}{c}{$I$} & 
\multicolumn{1}{c}{$F$} & 
\multicolumn{1}{c}{$I$} 
\\
\hline
[O~{\sc ii}] 3727                           &   2330  &   8360  &  27.7   &   38.3  \\
{[Ne~{\sc iii}]} 3869                       &    197  &     622 & $\dots$ & $\dots$ \\
H$\zeta$ + He~{\sc i} 3889                  & $\dots$ & $\dots$ &  17.1   &   28.2  \\
H$\epsilon$ + [Ne~{\sc iii}] 3969           & $\dots$ & $\dots$ &  14.8   &   37.8  \\
He~\sc{i} 4026                              & $\dots$ & $\dots$ &   4.1   &    5.0  \\
{[S~{\sc ii}]} 4069                         &   43.4  &     111 & $\dots$ & $\dots$ \\
H$\delta$ 4101                              & $\dots$ & $\dots$ &  22.5   &   34.2  \\
H$\gamma$ 4340                              & $\dots$ & $\dots$ &  46.7   &   62.8  \\
{[O~\sc{iii}]} 4363                         &   16.1  &   30.6  & $\dots$ & $\dots$ \\
He~\sc{i} 4471                              & $\dots$ & $\dots$ &   8.3   &    9.9  \\
C~{\sc i}:O~{\sc ii}:N~{\sc ii}:4562+4570   &   27.6  &   39.6  & $\dots$ & $\dots$ \\ 
He~\sc{ii} 4686                             &    108  &    134  & $\dots$ & $\dots$ \\
H$\beta$ 4861                               &    100  &    100  &     100 &     100 \\
{[O~\sc{iii}]} 4959                         &    113  &    101  & $\dots$ & $\dots$ \\
{[O~\sc{iii}]} 5007                         &    386  &    324  & $\dots$ & $\dots$ \\
He~{\sc i} 5016                             & $\dots$ & $\dots$ &    3.5  &    2.8  \\
{[N~{\sc i}]} 5199                          &    114  &   75.6  &    4.7  &    3.2  \\
{[N~{\sc ii}]} 5755                         &    159  &   60.2  & $\dots$ & $\dots$ \\
He~{\sc i} 5876                             &   34.1  &   20.5  &   39.7  &   19.3  \\
{[O~{\sc i}]} 6300                          &    339  &   95.5  & $\dots$ & $\dots$ \\
{[S~\sc{iii}]} 6312                         &   18.5  &    4.5  & $\dots$ & $\dots$ \\
{[O~{\sc i}]} 6363                          &    121  &   28.3  & $\dots$ & $\dots$ \\
{[N~\sc{ii}]} 6548                          &   1590  &    330  &   50.0  &   21.1  \\
H$\alpha$ 6563                              &   1460  &    274  &    666  &    301  \\
{[N~\sc{ii}]} 6584                          &   6820  &   1390  &    214  &     90  \\ 
He~{\sc i} 6678                             & $\dots$ & $\dots$ &    10.0 &    4.0  \\
{[S~\sc{ii}]} 6717                          &    727  &    135  &   42.0  &   16.7  \\
{[S~\sc{ii}]} 6731                          &    611  &    112  &   32.4  &   12.8  \\
He~{\sc i} 7065                             &   $\dots$ & $\dots$  &    6.6  &    2.1  \\
C~{\sc ii} 7064 + C~{\sc i} 7066            &    8.4  &    1.2  &   $\dots$ &  $\dots$ \\
{[Ar~\sc{iii}]} 7135                        &   80.2  &   11.1  & $\dots$ & $\dots$ \\
C~{\sc ii} 7236                             &   33.1  &    4.3  & $\dots$ & $\dots$ \\
C~{\sc i}:O~{\sc ii} 7289                   &   25.5  &    3.2  & $\dots$ & $\dots$ \\
{[O~\sc{ii}]} 7320                          &    254  &   31.6  & $\dots$ & $\dots$ \\
{[O~\sc{ii}]} 7330                          &    212  &   26.2  & $\dots$ & $\dots$ \\
N~{\sc i}:O~{\sc ii}:7379                   &   22.6  &    2.7  & $\dots$ & $\dots$ \\
\hline
$\log F$(H$\beta$) (\uflux)                 & $-$14.9  &        & $-$13.3  &        \\
$c$(H$\beta$)                               & 2.16$\pm$0.21 &   & 1.09$\pm$0.11 &   \\
\hline
\end{tabular}
\end{table}

The observed line flux ratios $F$ with respect to an arbitrary intensity of 100 for the \hb line together with the \hb flux for the inner and outer shells of HuBi\,1 are presented in Table~\ref{tbl:t_f_OSI}. 
The features at $\simeq$4566 \AA, $\simeq$7236 \AA, $\simeq$7289 \AA, and $\simeq$7379 \AA\ seem stellar, but none of them except the C~{\sc ii} $\lambda$7236 line can be unambiguously identified. 
It shall be noted that the emission arising from the inner shell for a number of emission lines is strongly contaminated by their emission arising from the outer shell.  
This issue is aggravated for the OSIRIS data, whose low spectral resolution does not allow the spectral split of the emissions from the inner and outer shells.  
Following the PI, PE, IwE and EwI classification of the different emission lines according to the relative contribution of the emission of the outer shell to that of the inner shell provided in Section 3.2, the flux of each line has been determined as described in Appendix~\ref{sec:OSIRIS_data}.

The intensities of the \ha and \hb lines have been used to derive the value of the logarithmic extinction coefficient, $c$(\hbs), for the inner and outer shells of HuBi\,1. 
A recombination Case B was adopted with values for the theoretical I(\has)/I(\hbs) lines flux ratio \citep{Osterbrock2006} according to the physical conditions of each shell following \citet{UO2021}'s prescriptions. 
Thus a theoretical I(\has)/I(\hbs) value of 2.74 was adopted for the inner shell, suitable for a value of $T_{\rm e}$ of 20000~K (close to that reported in Section~\ref{sec:phys_cond}), and 3.04 for the outer shell, suitable for a value of $T_{\rm e}$ of 5,000~K \citep{Guerrero2018,Pena2021}.  
The values of $c$(\hbs) derived for the inner and outer shells are 2.16$\pm$0.32 and 1.09$\pm$0.16, respectively. 
These have been used in conjunction with the extinction curve of \cite{Howarth1983} to deredden the relative line intensity ratios $I$ also presented in Table~\ref{tbl:t_f_OSI}. 

\section{Results}

The GTC MEGARA and OSIRIS data presented above allow a detailed spatially resolved spectroscopic investigation of HuBi\,1.  
The results of these analyses are provided in the next sections.

\subsection{Extinction maps}
\label{sec:extincion}

The determination of \cbe in the previous section reveals notable higher extinction values for the inner shell. 
This result is in sharp contrast with the results presented by \citet[][]{Pollacco} and \citet{Pena2021} that did not find spatial variations of the extinction across HuBi\,1, with similar values of $c$(\hbs) for the inner and outer shells. 
These studies did not accounted for the contamination of the emission of the inner shell by the foreground emission of the outer shell, which is particularly severe for the \ha and \hb Balmer lines used for the determination of $c$(\hbs).


The tomographic capability of the GTC MEGARA observations highlighted in Sections~\ref{sec:isolation} and \ref{sec:isolation-gaussian} to isolate the \ha and \hb emissions of the inner shell from those of the outer shell can be used to investigate the details of the spatial variations of $c$(\hbs) in HuBi\,1.  
As described in Section~\ref{sec:isolation-gaussian}, the maps derived using multi-Gaussian fits on a spaxel-by-spaxel basis (lower panels of Fig.~\ref{fig:maps_int}) are affected by small variations of the fit parameters and are noisier than the maps derived isolating the wings of the high-velocity components (upper panels of Fig.~\ref{fig:maps_int}). 
The maps derived using the latter method will be used to produce cleaner $c$(\hbs) maps, although the velocity range of the \ha and \hb emission profiles will be restricted to be the same for both lines, as allowed by the SNR of the fainter \hb line: 
from $-$14 \ve\ to $+$13 \ve\ for the approaching component and from $+$115 \ve\ to $+$122 \ve\ for the receding component. 
In this way, it is warrantied that the \ha to \hb ratio are derived from similar velocity ranges, although the velocity range for \ha is narrower than that used for the images presented in Figure~\ref{fig:maps_int}.

\begin{figure*}
\includegraphics[width=0.7\linewidth]{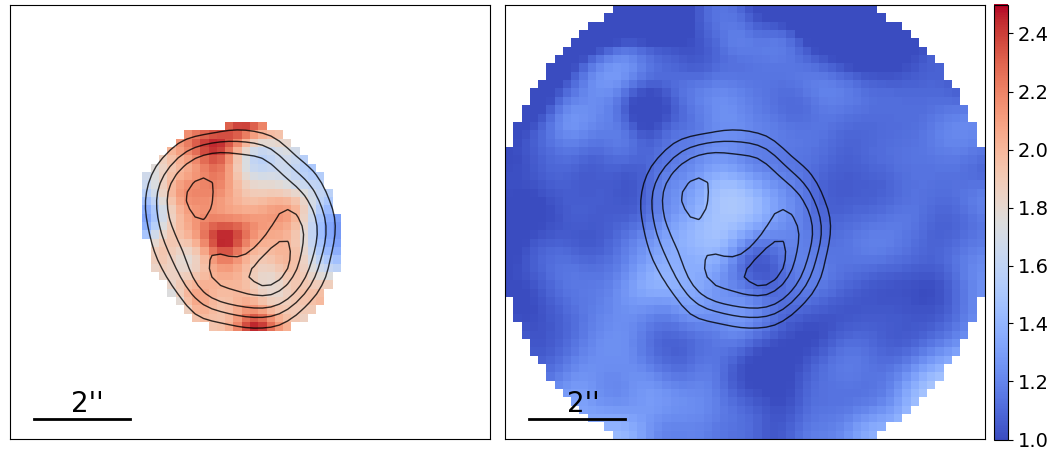}
\caption{
GTC MEGARA maps of the logarithmic extinction coefficient \cbe of the inner (left) and outer (right) shells of HuBi\,1 derived from the H$\alpha$ to H$\beta$ line flux ratios. 
Contours of the inner shell derived from the [O~{\sc iii}] map are overlaid on the maps. 
Pixels at the corner regions in the map of the outer shell have been ignored due to vignetting effects. 
}
\label{fig:cbeta}
\end{figure*}

The final extinction maps of the inner and outer shells of HuBi\,1 derived from the \ha and \hb flux map ratios are presented in Figure~\ref{fig:cbeta}. 
These maps indeed confirm that the extinction in the inner shell is larger than that of the outer shell.  
In order to compare the values in these \cbe maps with the values of \cbe derived in the previous section, average values of \cbe have been derived for the inner and outer shells. 
For this calculation, pixels with SNR $\leq$3 have been ignored.  
In addition, pixels of the inner shell beyond the second contour in Figure~\ref{fig:cbeta}-\emph{left} were excised because the low \hb flux, as well as pixels of the outer shell at the corners of the MEGARA IFU FoV where vignetting effects are appreciable.  
The averaged values of \cbe derived for the inner and outer shells are 2.09$\pm$0.21 and 1.14$\pm$0.11, respectively, which are consistent with the values derived in the previous section.  \\

The extinction maps presented in Figure~\ref{fig:cbeta} and the averaged values derived above confirm that the extinction in the inner shell is noticeable higher than that of the outer shell. 
The spatial distribution of the extinction in the inner shell revealed in the left panel of Figure~\ref{fig:cbeta} seems to peak in the innermost regions of the inner shell, at a location between the two peaks of the [O~{\sc iii}] map where the CSPN IRAS\,17154$-$1555 is located.
The extinction in the outer shell is basically flat, but a small increase is found in the regions of the outer shell projected onto the inner shell.  
This can be attributed to the absorption of the emission from the receding outer shell as it goes through the inner regions of HuBi\,1.

\citet{Guerrero2018} attributed the brightness decrease of the CSPN of HuBi\,1 of $\simeq$10 mags in the last 50 years to the ejection of C-rich material that  condensed into dust grains and veiled the light from the star rather than to surface temperature of the CSPN.  
The larger extinction of the inner shell of HuBi\,1 found here, 
in conjunction with the wealth of carbon emission lines detected in its 1991 spectrum \citep{Pollacco},
lends further support to the hypothesis that it contains C-rich material from a recent ejecta, while the outer shell is formed by pristine nebular material. 



\subsection{Diagnostic diagrams}
\label{sec:BPT}

The peculiar ionisation structure of HuBi\,1, with emission from forbidden lines of [O~{\sc iii}] and [N~{\sc ii}] dominating the inner shell, but emission from recombination lines of H~{\sc i} and He~{\sc i} in the outer shell, and the presence of He~{\sc ii} emission in the inner shell, which cannot be photoionized by its CSPN as its effective temperature is $T_\mathrm{eff}\approx$38,000~K, led \citet{Guerrero2018} to propose that the inner shell is mostly shock-excited.
The association of the inner shell of HuBi\,1 with a fast expanding ($\simeq$300~km~s$^{-1}$) shell-like structure ejected around 200 yrs ago \citep{Rechy} provided the means for shock-excitation. 
More recently, \citet{Pena2021} attributed temporal variations in different line intensity ratios of the inner shell and the increase of the electron temperature in this shell to shock-excitation, but the limitations of their analysis of the spectrum of the inner shell pose questions on their reliability. 
In particular, the line ratios presented by these authors based on the same GTC OSIRIS data set have not been corrected from DCR effects, which questions the reported line intensity ratio variations (see Appendix~\ref{sec:chromatic}).


Line intensity ratios diagrams have long been used to distinguish between different ionisation mechanisms \citep[for instance, the ones defined by][]{BPT}. 
These diagrams, often used to classify galaxies among AGNs, starburst galaxies or LINERs \citep{VO1987,Kewley2001}, can also be applied to PNe.
The usual line intensity ratios used for line ratio diagrams are 
[O~{\sc iii}]/\hb vs [N~{\sc ii}]/\has, [S~{\sc ii}]/\ha or [O~{\sc i}]/\has, but other ratios such as 
[O~{\sc iii}]/\ha vs [S~{\sc ii}]/\ha or [N~{\sc ii}]/\ha  have been used in proto-planetary nebula (pPNe) and PNe and their FLIERs as well \citep[e.g.,][]{Raga_BPT}. 

\begin{figure*}
\centering
\includegraphics[width=0.8\linewidth]{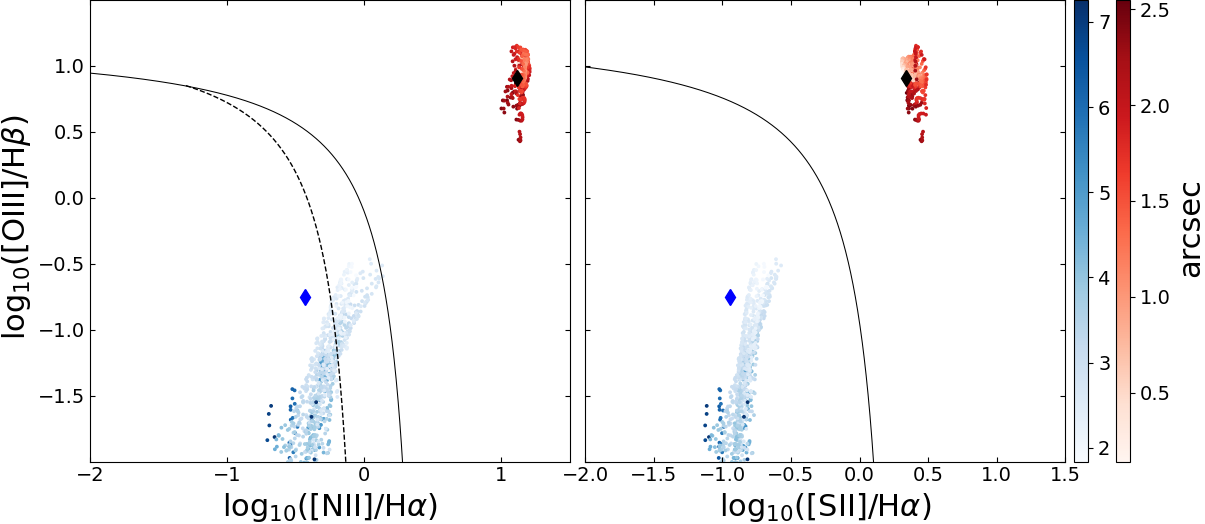}
\caption{
Pixel by pixel distributions of the [O~{\sc iii}]$\lambda5007$/\hbs, [N~{\sc ii}]\,$\lambda 6584$/\has, and [S~{\sc ii}]$\lambda(6716 + 6730)$/\ha line intensity ratios in the corresponding BPT diagrams for the inner (red points) and outer (blue points) shells of HuBi\,1.  
The blue scale represents the distance of the points of the outer shell to the centre of the nebula from 2\farcs5 to 7\farcs5, whereas the red scale shows the distance of the points of the inner shell to the centre of the nebula from 0 to 2\farcs5.
Data points representative of the inner and outer shells of HuBi\,1 derived from the GTC OSIRIS long-slit spectra are shown as cyan and magenta diamonds for the outer and inner shell respectively. 
The solid and dashed curves correspond to the theoretical models of \citet{Kewley2006} and \citet{Kauffmann2003}, respectively. 
}
\label{fig:bpt}
\end{figure*}

The GTC MEGARA \hbs, \has, [O~{\sc iii}], [N~{\sc ii}], and [S~{\sc ii}] maps of the inner and outer shells of HuBi\,1 have been used to obtain the pixel-by-pixel distributions of the log$_{10}$([O~{\sc iii}]/\hbs) vs log$_{10}$([N~{\sc ii}]/\has) and log$_{10}$([S~{\sc ii}]/\has) in the line ratio diagrams presented in Figure~\ref{fig:bpt}. 
Integrated ratios for the inner and outer shell have been also computed using the GTC OSIRIS long-slit data.
The extinction effects in these line ratios is expected to be negligible, given the small wavelength difference, yet the intensity ratios of all points have been corrected using the extinction maps in Figure~\ref{fig:cbeta}. 

The line ratio diagrams in Figure~\ref{fig:bpt} clearly reveal that two different mechanism of ionisation are present in HuBi\,1, being the inner shell dominated by shocks while the outer shell is undoubtedly photoionized. 
We note that the interpretation of diagnostic diagrams using spatially-resolved IFU data can be misleading due to ionization stratification mimicking the behaviour of shock-excited nebulae \citep{M2018}, but this is not the case here as the integrated values of the line ratios obtained from GTC OSIRIS long-slit data share similar loci in these diagrams (Fig.~\ref{fig:bpt}).
The spread of the data points in these diagrams is notably larger than that in the line ratio diagrams presented by \citet{Pena2021}, with values of log$_{10}$([O~{\sc iii}]/\hbs), log$_{10}$([N~{\sc ii}]/\has) and log$_{10}$([S~{\sc ii}]/\has) close to unity for the inner shell. 
These points are located in the region of these line ratio diagrams that can be interpreted as the result of shock-excitation. 
The spectral coverage of our MEGARA data does not cover the shock-sensitive [O~{\sc i}] $\lambda 6300$ emission line, but we note that the value of the [O~{\sc i}] $\lambda 6300$/\ha line intensity ratio derived from the OSIRIS data for the inner shell, $\simeq$0.95, locates it in the region of shock excitation of the log$_{10}$([O~{\sc iii}]$\lambda 5007$/\hbs) vs log$_{10}$([O~{\sc i}] $\lambda 6300$/\has) line ratio diagram.

The distributions of the points of the outer shell in these line ratio diagrams show a notable spread, with the innermost points closer to the shock ionisation region in the line ratio diagrams and the more external points well located in the photoionisation region of these diagrams. 
Such correlation between ionisation and radial distance could originate from the regions of diffuse [S~{\sc ii}] and [N~{\sc ii}] in the so-called intermediate shell, whose ionisation may include the effects of shocks. 
The born-again 3D radiation-hydrodynamic model of HuBi\,1 presented by \citet{Jesus} suggests that these structures can appear as the result of the diffusion of turbulence from the ejecta onto the outer shell, which would explain the mixed excitation.  

\subsection{Physical conditions of the inner shell}
\label{sec:phys_cond}

\begin{figure}
\centering
\includegraphics[width=1.0\columnwidth]{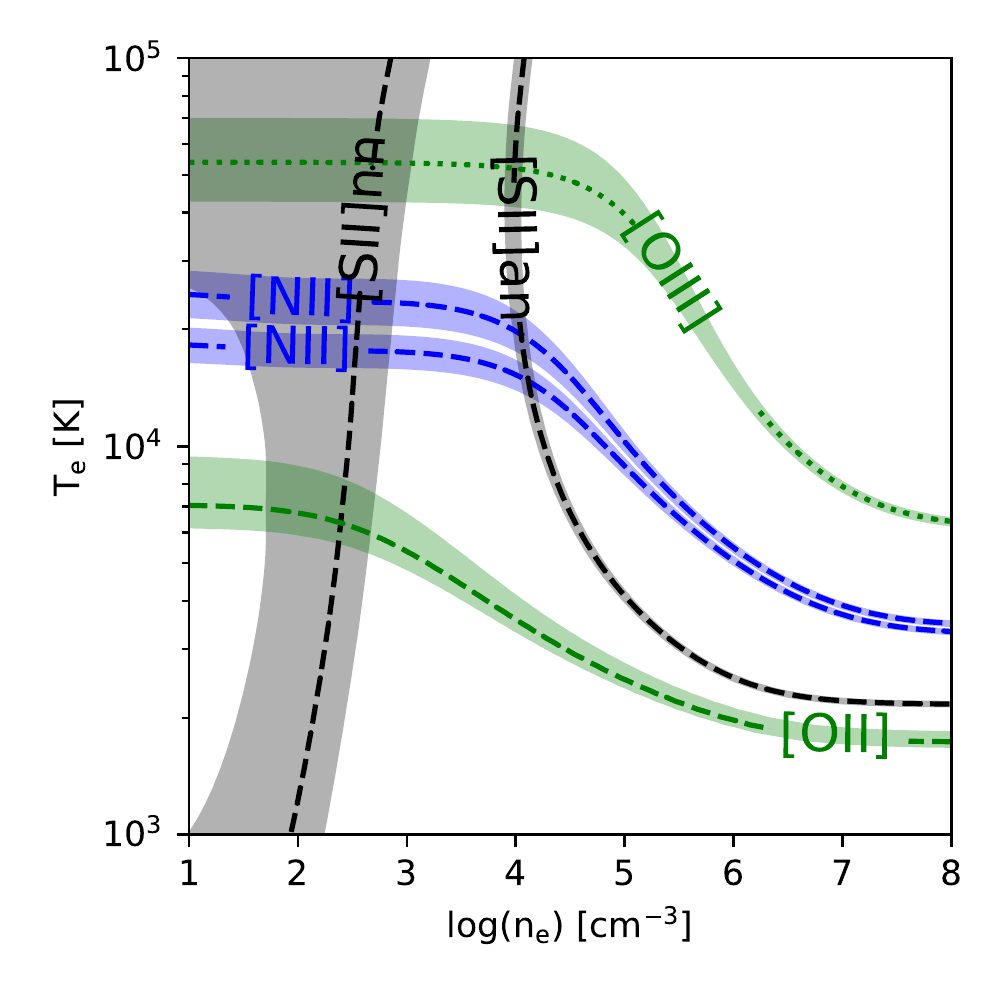}
\caption{
Diagnostic diagrams for the different temperature- and density-sensitive line ratios available in the spectrum of the inner shell of HuBi\,1. 
The shaded areas have been computed assuming a 10\% uncertainty for the measured line intensities. 
The `an' and `nn' labels for the [S~{\sc ii}] line ratios stand for the auroral $\lambda$4069 to nebular $\lambda\lambda$6716,6731 and nebular $\lambda$6716 to nebular $\lambda$6731 line ratios, respectively. 
}
\label{fig:NeTediag}
\end{figure}

We used PyNeb \citep{LMS2015} to investigate the physical conditions of the inner shell\footnote{
An investigation of the physical conditions of the outer shell cannot be performed because no temperature-sensitive auroral lines are detected in its spectrum.
} of HuBi\,1. 
The corresponding diagnostic diagrams are shown in Figure \ref{fig:NeTediag}, where the shaded areas have been computed assuming a 10\% uncertainty for the measured line intensities.

As for the electron density, $n_{\rm e}$, the [S~{\sc ii}] $\lambda$6716 to $\lambda$6731 line ratio (the label `nn' stands for nebular-nebular) draws an area for $n_{\rm e}$ ranging from a few tens to 1000 cm$^{-3}$. 
This value is constrained to 200-600 cm$^{-3}$ in regions where the density diagnostic curve for the [S~{\sc ii}] $\lambda$6716 to $\lambda$6731 line ratio crosses the temperature diagnostic curves.

As for the electron temperature, $T_{\rm e}$, the situation is a bit more complex: 
the [O~{\sc iii}] $\lambda$4363 to $\lambda$5007 line ratio points to $T_{\rm e}$ between 45,000 K and 70,000~K, while the [N~{\sc ii}] $\lambda$5755 to $\lambda\lambda$6548,6584 line ratio points to much lower values, in the range from 16,000~K to 26,000~K. 
This result can be interpreted as lower ionization N$^+$ species prevails over higher ionization O$^{++}$ species in regions where the gas have had more time to recombine since the shock passed, and then to cool down. 
The situation is even more extreme for the [O~{\sc ii}] $\lambda$3727 to $\lambda\lambda$7320,7330 ratio, which implies a very cold T$_\mathrm{e}$, in the range between 6,000~K and 9,000~K, and for [S~{\sc ii}] $\lambda\lambda$6716,6731 to $\lambda$4069 line ratio (labeled `an' for auroral-nebular), which points to a very high T$\mathrm{e}$, not even shown in the diagram. 
These results would be hard to reconcile in a photoionized gas, since the O$^+$, N$^+$, and S$^+$ regions are expected to be co-spatial, thus sharing similar physical properties.  
In the case of shocked regions, however, the steep $T_{\rm e}$ gradient may cause the emission of a given species to arise from regions at very different temperatures.

The extreme values of $T_{\rm e}$ derived from the [O~{\sc ii}] and [S~{\sc ii}] temperature-sensitive line ratios may have an alternative explanation. 
Both line ratios involve red and blue lines, but in the opposite way, with the nebular [O~{\sc ii}] $\lambda$3727 line and the auroral [S~{\sc ii}] $\lambda$4069 line in the blue range.  
In the next sections, we will see that the blue [O~{\sc ii}] $\lambda$3727 and [S~{\sc ii}] $\lambda$4069 lines, but also the [Ne~{\sc iii}] $\lambda$3869 line and in a less extent the [O~{\sc iii}] $\lambda$4363 line are systematically underpredicted by shock models. 
If the intensity of the [O~{\sc ii}] $\lambda$3727 line and [S~{\sc ii}] $\lambda$4069 lines was, indeed, overestimated, it would result in the predicted low $T_{\rm e}$ for the [O~{\sc ii}] line ratio and high $T_{\rm e}$ for the [S~{\sc ii}] line ratio in Figure~\ref{fig:NeTediag}.  
Indeed, the value of $T_{\rm e}$ derived from the [N~{\sc ii}] line ratio is the most insensitive to extinction, as it involves lines in the red spectral range.

\subsection{Electron density map}\label{sec:temp_dens}

The spatial distribution of the nebular density can be derived from the density-sensitive [S~{\sc ii}] $\lambda6716,6731$ doublet intensity ratio independently of the excitation mechanism (shocks or photoionisation). 
As described in Section~\ref{fig:total_flux} and Appendix~\ref{sec:OSIRIS_data}, the [S~{\sc ii}] emission lines are classified in Table~\ref{tbl:shift_cla} as IwE, i.e., the emission is mostly attributed to the inner shell with some contribution at systemic velocities arising from the intermediate shell. 
Their flux maps have thus been computed using a method similar to that applied to the \ha and \hb lines in Section~\ref{sec:isolation} with the small differences detailed in Appendix~\ref{sec:OSIRIS_data}.

\begin{figure}
\centering
\includegraphics[width=0.9\columnwidth]{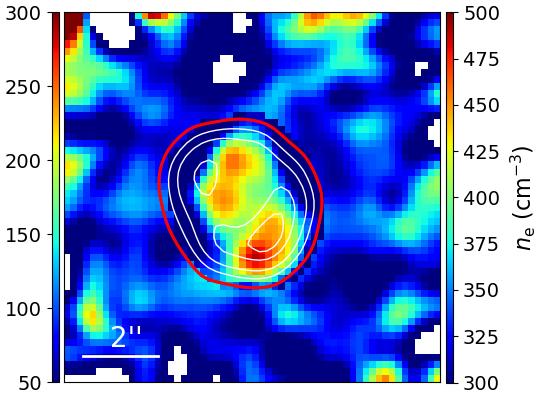}
\caption{
Electron density $n_\mathrm{e}$ map of HuBi\,1 derived from the [S~{\sc ii}] $\lambda6716$ to [S~{\sc ii}] $\lambda6730$ line intensity ratio. 
The contours of the emission in the [O~{\sc iii}] $\lambda$5007 line showing the extent of the inner shell are overlaid. 
The most extended red contour separates the inner and outer shells of HuBi\,1. 
Their densities $ n_\mathrm{e}$ were computed assuming an electron temperature of 30,000~K for the former (right colorbar) and 5,000~K for the latter (left colorbar). 
}
\label{fig:nmap}
\end{figure}

The resulting [S~{\sc ii}] $\lambda\lambda$6716,6730 maps are then used to obtain the electron density map using the usual relation between the [S~{\sc ii}]$\lambda6716$/[S~{\sc ii}]$\lambda6730$ ratio and $n_\mathrm{e}$ \citep{Osterbrock2006} 
assuming for $T_\mathrm{e}$ a value of 30,000~K for the inner shell, within of range of values derived from the [N~{\sc ii}] and [O~{\sc iii}] temperature-sensitive ratios in the previous section, and 5,000 K for the outer shell.
Figure~\ref{fig:nmap} shows the electron density map for all pixels with SNR$>3$. 
We obtained averaged densities of f$n_\mathrm{e} \approx$500~cm$^{-3}$ for the inner shell and $n_\mathrm{e} \approx$100~cm$^{-3}$ for the outer shell, the latter in the low-density limit of the density diagnostic provided by the [S~{\sc ii}] doublet. 
The electron density of the outer shell is quite flat, but the inner shell shows two peaks more or less coincident with the [O~{\sc iii}] emission peaks.

The electron density of the inner shell agrees with that reported by \citet{Guerrero2018}, whereas the extreme high electron density $n_\mathrm{e} \simeq$20,000 cm$^{-3}$ reported by \cite{Pollacco} can be discarded as those authors did not account for the contamination of the [S~{\sc ii}] emission lines by stellar C~{\sc ii} lines. 
\citet{Pena2021} reports values for the electron density of the inner and outer shells of HuBi\,1 for different epochs between 2000 and 2018 that are suggestive of higher $n_\mathrm{e}$ for the inner shell than the outer shell, although the large spread of $n_\mathrm{e}$ values makes such result only qualitatively consistent with those obtained here.  

\subsection{Chemical abundances}\label{sec:abundancias}

Using diagnostic diagrams and the scheme laid out by BPT, we have demonstrated in Section~\ref{sec:BPT} that shocks is the main excitation mechanism of the inner shell of HuBi\,1.  
Indeed, the high values of the auroral-to-nebular line ratios [O\,{\sc iii}] $\lambda$4363 \AA/[O\,{\sc iii}] $\lambda$5007 \AA\ $\sim$ 0.09 and [N\,{\sc ii}] $\lambda$5755 \AA/[N\,{\sc ii}] $\lambda$6584 \AA\ $\sim$0.33 are also clear indicators of shocks \citep{Dopita_1977}. 
Since shock excitation has to be taken into account for abundances determination in the inner shell of HuBi\,1, we computed those making use of the {\sc mappings v} code \citep{Sutherland2018}. 
Details of the code can be found in Appendix~\ref{sec:app_mappings}.





For our modelling, we consider shocks as the only source of ionisation in the inner shell as the large extinction towards the CSPN basically absorbs all its UV emission. 
We ran a number of models taking into account the emission from both the shock and its precursor \citep{Allen_2008, Sutherland_2017}. 
According to analyses of spectroscopic observations of the outer shell of HuBi\,1 \citep[e.g.,][]{Guerrero2018}, shocks in these simulations are assumed to propagate into a fully ionised medium (i.e., H/H$^{+}$ = He/He$^{+}$ = 1) with a preshock temperature of 10$^{4}$~K.

We start by fixing the values of the shock velocity $v_\mathrm{s}$ at 80~km~s$^{-1}$ as smaller velocities can be precluded \citep{Rechy}, the pre-shock density $n$ at 1~cm$^{-3}$ \citep[similar to that found for non-fully radiative young shocks by][]{Dopita_2018}, and the strength of the magnetic field $B$ at 1~$\mu$G. 
The later is low for a PN \citep{Rodriguez_2017}, but plausible if we assume that magnetic field scales with density $B \propto n^{1/2}$ as proposed by \citet{Cox_2005} for average ISM properties.

We then ran a number of {\sc mappings v} models varying the O and N abundances, keeping the He abundance at 12+log$_{10}$(He/H)=13.0 and scaling the abundances of the other heavy elements with those of O according to the Solar abundances \citep{Asplund_2009}. 
The {\sc mappings v} models studying the variations of the O and N abundances are presented in Table~\ref{tab:c1} of  Appendix~\ref{sec:app_mappings}, where the model line emission strengths are compared with those observed reported in Table~\ref{tbl:t_f_OSI} to determine the best values for the O and N abundances. 
The emission line ratio [O\,{\sc iii}] $\lambda$4363 \AA/[O\,{\sc iii}] $\lambda$5007 \AA, which is particularly sensitive to the shock conditions \citep{Dopita_1977}, is used as a main indicator, but the intensities of the [O\,{\sc iii}], [O\,{\sc ii}] and [N\,{\sc ii}] emission lines are also considered

The N/H abundances were first fixed to their Solar values and the O/H abundances allowed to vary (columns 3 to 7 of Tab.~\ref{tab:c1}).
It can be seen there that, although O abundances below 12+log$_{10}$(O/H)$<$9.8 reproduce best the auroral-to-nebular [O\,{\sc iii}] ratio, they clearly underestimate the flux of all O emission lines. 
Higher O abundances tend to overestimate the [O\,{\sc iii}] emission lines and predict auroral-to-nebular [O\,{\sc iii}] ratios much smaller than observed. 
The model with 12+log$_{10}$(O/H)=9.8$\pm$0.1, which best reproduces most of the [O\,{\sc iii}] emission lines, is adopted as best fit model. 
The O/H abundances were then fixed to this value and the N/H abundances allowed
to vary (columns 8 to 12 of Tab. C1). 
As for the O abundances, the intensity of the N lines require abundances several orders of magnitudes higher than Solar values, with the best fit achieved at 12+log$_{10}$(N/H)=$9.9\pm0.10$.

We tried then to assess the He abundance that best reproduces the intensity of the He recombination lines. 
Table~\ref{tab:c2} shows that predicted intensity of the He recombination lines scales with the He abundance, although it also has noticeable effects in important collisional lines, such as [N\,{\sc ii}] $\lambda$6584, implying variations at factors of about one half in the N abundances anti-correlated with those of He.  
On the other hand, the intensity of the O lines and thus their abundances are quite insensitive to variations in the He abundances. 
We varied the He abundance around a value of 12+log$_{10}$(He/H)=13, but the He\,{\sc ii} $\lambda$4686 line was always underestimated, whereas the He\,{\sc i} $\lambda$5876 line was clearly overestimated. 
A value of 12+log$_{10}$(He/H)=13.00$\pm$0.15 was selected as a best compromise.  

Finally, we investigated whether the adopted pre-shock density and shock velocity may explain the discrepancies found in some emission line ratios for our fixed set of chemical abundances. 
In  Table~\ref{tab:c3} we show that increasing the shock velocity $v_\mathrm{s}$ from 100 to 200~km~s$^{-1}$ tend to decrease all emission line ratios with respect to the [O~{\sc iii}] emission lines.
Therefore, we conclude that a shock velocity $v_\mathrm{s}$ of 80~km~s$^{-1}$ is the best choice in agreement with \citet{Guerrero2018}. 
Similarly, we show in Table~\ref{tab:c4} that increasing the pre-shock density makes closer the estimation of the [O\,{\sc iii}] ratio to the observed one, but other important emission lines as the [S\,{\sc ii}] doublet, [O\,{\sc ii}] and [N\,{\sc ii}] lines are not well reproduced. 
Lower densities lead to lower [O\,{\sc iii}] $\lambda$4363/[O\,{\sc iii}] $\lambda$5007 ratios, making them differ more notably from the observed one. 
Therefore, the best models are adopted to have a pre-density  $n_\mathrm{pre}$ of 1~cm$^{-3}$.

We list in Table~\ref{tab:model} the parameters of our preferred model to reproduce the shock-excited spectrum of the inner shell of HuBi\,1. 
We note that the C abundances in this table are scaled from the O abundances assuming a Solar C/O ratio of $\sim$0.55 \citep{Asplund_2009}, as the available line ratios do not provide a suitable constraint to the C abundances.   
The [O\,{\sc iii}] and [N\,{\sc ii}] lines are fairly well described by this model, with the most notable discrepancies for the He~{\sc i} $\lambda$5876 and [O\,{\sc ii}] $\lambda$3727 emission lines. 
These discrepancies will be discussed in further details in the next section.


\begin{table}
\caption{{\sc mappings v} model considered to reproduce the emission from the inner shell in HuBi\,1. The emission lines are referred to $I$(H$\beta$)=100.}
\setlength{\tabcolsep}{\tabcolsep}    
\label{tab:model}
\begin{tabular}{lccc}
\hline
Parameter & Preshock & Postshock \\ 
\hline
Density ($n$)[cm$^{-3}$] & 1  & 3.7 \\
Electronic Density ($n_\mathrm{e}$)[cm$^{-3}$] & 5.8  & 21.4 \\
Temperature ($T$)[K] & 30000  & 380000  \\ 
Magnetic field ($B$)[$\mu$G] & 1 & 3.7 \\ 
\hline
Chemical abundances & & 12+log$_{10}$(X/H) \\ 
\hline
H                   & & 12.00              \\
He                  & & 13.00$\pm$0.15    \\
C                   & & 9.53      \\
N                   & & 9.90$\pm$0.10     \\
O                   & & 9.80$\pm$0.10      \\
Ne                  & & 9.03          \\
\hline
Line          & Observation & Model \\ 
\hline   
\oii~3727         &   8360  & 571.2   \\
{[Ne~\sc{ iii}]}~3869       &  622    & 54.9 \\
\sii~4069         &  111    & 14.2 \\
\oiii~4363        &  30.6   & 25.0 \\ 
He\,{\sc ii} 4686 &   134   & 66.8 \\ 
\oii~4959         &   101   & 129.2 \\
\oiii~5007        &   324   &  373.4 \\
\nii~5199         &  75.6   &  119.3 \\
\nii~5755         &   60.2  &  66.3 \\
He\,{\sc i} 5876  & <20.5   & 336.7 \\
\oi~6300          & 95.5    & 30.4 \\
{[S~\sc{ iii}]}~6312        &   4.5   & 5.3  \\
\oiii 6363        & 28.3    & 9.7 \\
\nii~6548         &  330    & 552.2 \\
\ha               & 274     & 407.4 \\
\nii~6584         &   1390  &   1624 \\
\sii~6717         &   135   & 104.0 \\
\sii~6731         &   112   & 115.0 \\
{[Ar~{\sc iii}]}  & 11.1     & 7.8\\
\oii~7320         & 31.6    & 43.1 \\
\oii~7330         & 26.2    & 34.8 \\ 
\hline
\end{tabular}
\end{table}

\section{Discussion}

HuBi\,1, the inside-out PN, has proven to be a remarkable object, with extreme changes recorded in the past decades \citep[e.g.,][]{Guerrero2018,Pena2021}. 
Similar variations in human-time scales have also been reported for the two youngest born-again PNe identified thus far, namely the Sakurai's Object and A\,58, a.k.a.\ Nova Aql 1919 \citep[see][and references therein]{Clayton2013,Evans2020}, as well as for the LTP star SAO\,244567 and the nebula around it \citep{Reindl2017,Balick2021}. 
Indeed, the dramatic decline in brightness of the CSPN of HuBi\,1 by $\simeq$10 mag in $\sim$50 years and the strong IR emission and the profusion of C emission lines from the innermost regions detected in the oldest available spectra \citep{Pollacco} are suggestive of the ejection of highly-enriched material that had resulted in the sudden formation of large amounts of dust, well in agreement with a born-again scenario.  
If this were the case, there are a number of predictions that can be uniquely tested using the unequalled capabilities of MEGARA at the GTC to disentangle the inner shell emission from that corresponding to the outer H-rich PN.

Since VLTP events result in the production of large amounts of dust, one basic expectation would be that the inner shell of HuBi\,1 should be more extincted than the outer shell.  
We have for the first time obtained clean extinction maps of the inner and outer shells by dissecting the different contributions of their emission to the H Balmer lines.  
These unpolluted extinction maps confirm that the inner shell exhibits larger $c$(H$\beta$) values than the outer nebula, in contrast with the similar values for both shells reported in previous works \citep[e.g.,][]{Pena2005,Pena2021}.  
Incidentally, these extinction maps allow us the proper reddening correction of the spectra extracted for the inner and outer shells using the appropriate value of $c$(H$\beta$) for each of them.

After experiencing a VLTP event, the CSPN is expected to go back for a short period of time to the locus of the H-R diagram occupied by AGB stars, reducing considerably its effective temperature and thus its ionisation photon flux.
As a result, photoionisation cannot be the main excitation mechanism of the emission arising from the innermost regions of young born-again PNe.  
Combining MEGARA and OSIRIS data sets, we have obtained reddening-corrected [O~{\sc iii}]/H$\beta$, [N~{\sc ii}]/H$\alpha$ and [S~{\sc ii}]/H$\alpha$ line ratios that have been used to produce standard diagnostic diagrams.  
The position in these diagrams of data points from the inner shell undoubtedly demonstrate that the main emission mechanism in the inner shell of HuBi\,1 is shocks, confirming the original proposition explored by \citet{Guerrero2018}. 

The most critical test of the born-again nature for the inner shell of HuBi\,1 would be a H-poor abundance.  
We have computed those from the line ratios measured in the MEGARA and OSIRIS data sets using the {\sc mappings v} code to conform to the shock-excitation of this region.  
The best-fit model (Tab.~\ref{tab:model}) is capable to reproduce most of the emission lines detected in our spectra, although the strength of the [O~{\sc ii}] $\lambda$3727 doublet is notoriously underestimated, whereas those of the He~{\sc ii} $\lambda$4686 and He~{\sc i} $\lambda$5876 emission lines cannot be simultaneously matched.  
The significant differences between the observed and predicted intensities of the [O~{\sc ii}] $\lambda$3727 and [O~{\sc ii}] $\lambda\lambda$7320,7330 emission lines may reveal the complexity of the physical conditions and possibly varying optical depth of the O$^+$-emitting region. 
Alternatively, these discrepancies can be partially attributed to the effects of the relatively small photoionization produced by the cool CSPN of HuBi\,1 \citep[$T_\mathrm{eff}\approx$35--38~kK;][]{LH1998,Guerrero2018}, which would raise the emissivity of the [O~{\sc ii}] $\lambda$3727 doublet and He~{\sc i} $\lambda$5876 emission line. 
More likely there is an over-correction of the extinction of the intensity of the emission lines in the blue end of the spectrum due either to spatially varying absorption or to the use of a generic extinction curve of the interstellar medium, but probably not suitable for the peculiar chemical abundances of the ejecta. 
Section~\ref{sec:phys_cond} already noted the discrepancies between the electron temperature derived from line ratios involving [O~{\sc ii}] and [S~{\sc ii}] ``blue'' and ``red'' lines.  
Indeed, the shock models here presented does not only underestimate the ``blue'' [O~{\sc ii}] $\lambda$3727 and [S~{\sc ii}] $\lambda$4069 lines, but also the [Ne~{\sc iii}] $\lambda$3869 emission line.

The best-fit model in Table~\ref{tab:model} implies the notable hydrogen deficiency in the innermost ejecta of HuBi\,1, with He/H $\simeq$10, O/H $\simeq$0.006, and N/H $\simeq$0.008, well above those typical for PNe.  
These results are in sharp contrast with those recently presented by \citet{Pena2021}, who claimed that the chemical composition of the inner and outer shells of HuBi\,1 were similar among them and typical of PNe.  
We note, however, a number of flaws in the analysis presented by \citet{Pena2021}: 
(1)  the extinction-correction of the outer shell was applied to the inner shell of HuBi\,1, which is inadequate and result in the incorrect determination of the physical conditions of the inner shell, 
(2) the notable DCR effects in the same OSIRIS long-slit spectra presented here (see Appendix \ref{sec:OSIRIS_data}) were not corrected, 
(3) photoionisation was assumed as the excitation mechanism for the calculation of the inner shell chemical abundances, although they had proven the prevalence of shocks as the main excitation mechanism, and 
(4) more importantly, the dominant contamination of the outer shell to the H$\beta$ and H$\alpha$ emission from the inner shell was not subtracted, artificially enhancing the hydrogen content of the inner shell.

The abundances by number obtained in our analysis considering shocks are in excellent agreement with the abundances determined from collisionally-excited emission lines (CELs) from the H-poor knots of born-again PNe \citep{Jacoby1983,Manchado1988,Guerrero1996,Wesson2003,Wesson2008}. 
Our He/H abundance of 10 is as well in the same range from 3.2 to 11.7 to that obtained for A\,30 and A\,58 \citep[see table~7 in][]{Wesson2008}. 
It can thus be concluded that the chemical abundances of the inner shell of HuBi\,1 are consistent with the H-poor nature of a born-again scenario.

The chemical abundances of HuBi\,1, as those derived for other born-again PNe using CELs, are far from the theoretical expectations from born-again scenarios \citep[e.g.,][]{Miller}.  
It is well-known that observed chemical abundances in born-again PNe only become consistent with those theoretically predicted when they are computed using optical recombination lines (ORLs), as 
the abundances discrepancy factor (ADF), i.e., the ratio between the chemical abundances derived from ORLs and CELs, is extremely high among born-again PNe, with values 
up to 700 for A\,30 \citep{Wesson2003} and $\simeq$90 for A\,58 \citep{Wesson2008}.  
For comparison, the average ADF\footnote{
See \url{https://www.nebulousresearch.org/adfs} for the most up to date list of ADFs in PNe.
of hydrogen-rich PNe is $\simeq$2 \citep[][]{Wesson2018}.
}
Unfortunately, the ORLs needed for the determination of the ADF in HuBi\,1 are not available in its spectrum.  
If an ADF of 100 were adopted, which seems reasonable for born-again PNe, total ``ADF-corrected'' abundances by mass of 0.01, 0.58, 0.06, 0.16, 0.15 and 0.04 for H, He, C, N, O and Ne, respectively, would be obtained. 

These ``ADF-corrected'' chemical abundances are in the range of the theoretical expectations from born-again scenarios considered above \citep{Miller}, but for the N/O ratio close to unity, which seems otherwise
typical of nova events \citep[see][]{Lau2011}.  
This scenario, however, is unlikely, since the slowest novae have velocities $\sim300-700$~\ve \citep[see,e.g.,][and references therein]{Santamaria2020}, whereas the bulk of material of the hydrogen-poor ejecta of HuBi\,1 expands at 80--100~km~s$^{-1}$.

The details of these ``ADF-corrected'' abundances do not completely agree with those estimated for the CSPN of HuBi\,1 using non-LTE atmosphere models \citep{LH1998,Guerrero2018}: 
the nebular abundances by mass of 0.58 for He and 0.15 for O differ from the stellar values of 0.33 and 0.10, respectively, 
the nebular N/O ratio is about 10 times larger than the stellar N/O ratio (0.1), and 
the nebular C abundances are about 8 times smaller than the stellar ones (0.5).  
On the other hand, the stellar abundances are perfectly matched by the theoretical predictions \citep{Guerrero2018}. 
We suspect that the differences between the ``ADF-corrected'' abundances of HuBi\,1 and the stellar abundances and theoretical predictions arise from a notable underestimation of the C abundances and in a minor degree of the O abundances. 
The C and Ne abundances obtained with our shock model have been estimated using their Solar ratios with respect to the O abundances, but the wealth of C~{\sc ii} and C~{\sc iii} lines detected by \citet{Pollacco} point to a significant overabundance of C. 
Moreover, large amounts of C are to be trapped into C-rich dust such as amorphous carbon species as shown in the born-again PNe A\,30 and A\,78 \citep{Jesus2021b}. 
In addition, C and O atoms might be trapped into CO molecules, which in the case of the born-again PN A\,58 amount to a mass $\simeq$10$^{-5} M_\odot$ \citep{Tafoya2022}. 
If those effects were to be taken into account, the C fraction by mass would increase notably, reducing that of the other elements, and the N/O ratio would be reduced.  
The total abundances by mass of He, C, N, and O would then become consistent with those of 0.33, 0.5, 0.01, and 0.10, respectively, estimated for the CSPN \citep{Guerrero2018}. 
The case of the C abundances of the born-again ejecta of HuBi\,1 will be addressed in a subsequent paper (Rodr\'{i}guez-Gonz\'{a}lez et al., in prep.) using IR observations to obtain a coherent set of chemical abundances of H, He, C, N, and O \citep[see][]{Jesus2021b}.  
Future studies that simultaneously model the gas, dust and molecules of born-again PNe will help strengthen our understanding of the evolutionary sequences of the progenitors of such unique objects.





\section{Conclusions}

We have analysed integral field MEGARA and long-slit OSIRIS optical spectroscopic data of HuBi\,1, the inside-out PN. 
The combined capabilities of these two GTC instruments have allowed us to study the ionisation structure of HuBi\,1 in unprecedented detail and to determine the true chemical abundances of the recent ejecta in its innermost regions. 
Our main findings can be summarised as follows:

\begin{itemize}

\item
The multiple shell structure of HuBi\,1 is clearly dissected, with an inner shell associated with the recent ejecta, an outer shell, and additional emission mainly in the low-ionization [N~{\sc ii}] and [S~{\sc ii}] lines from an intermediate region. 
The inverted ionization structure of the inner shell of HuBi\,1 is dramatically confirmed, with the emission in the He~{\sc ii} $\lambda$4686 line encompassing that in the [O~{\sc iii}] lines, which in addition surrounds that of the low-ionization [N~{\sc ii}] and [S~{\sc ii}] emission lines. 
The morphology of the intermediate region and its excitation are indicative of the interaction between the expanding ejecta of the inner shell and the old nebula.  
This intermediate region can be envisaged as the precursor of the petal-like structure clearly developed in the evolved born-again PNe A\,30 and A\,78.

\item 
The high-dispersion integral field spectroscopic data obtained with MEGARA have proven to be conclusive to separate for the first time the faint emission of the H Balmer lines from the inner shell from the bright emission of the outer shell.  
The emission in the H$\beta$ and H$\alpha$ lines from the inner shell of HuBi\,1 have been accurately obtained, allowing us to compute reliable relative line intensity ratios to determine the spatially varying extinction and excitation, and to compute the true chemical abundances of the inner shell. 

\item
The extinction coefficient c(\hbs) of the inner shell ($\simeq$2.2) is on average twice that of the outer shell ($\simeq$1.1).  
The increased absorption towards the central regions of HuBi\,1 is consistent with the idea that large amounts of dust have been recently produced there as suggested by the remarkable brightness reduction of its CSPN, IRAS\,17514$-$1555, by $\sim$10 magnitudes in the last 50 years.

\item
The improved relative line intensity ratios have allowed us to undoubtedly confirm that the inner shell is ionised by shocks using BPT diagrams.  
On the other hand, the outer shell is located on the photoionisation zone of these diagram. 
Interestingly, the distributions of data points of the outer shell in the line ratio diagrams show a gradient, with points from the outermost regions clearly in the photoionisation zone, but points from the innermost regions of the outer shell spatially consistent with the intermediate region described above being closer to the locus for shocks. 
This supports the idea that the intermediate region is partially shock-excited by the interaction of the recent ejecta with the old nebula.

\item 
The total abundances by number considering shocks are in agreement with the chemical abundances obtained from CELs in the hydrogen-poor knots of other born-again PNe such as A\,30, A\,58, and A\,78. 
This is also the case for the total abundances by mass computed assuming an ADF of 100 similar to other born-again PNe.
In particular, the large amount of He complies the hydrogen-poor nature of the ejecta predicted in born-again scenarios. 
Our observation did not detect any carbon lines and we could not measure its abundances, but it can be expected that large amounts of carbon are to be trapped into C-rich dust.  
Moreover some carbon and oxygen can be found forming CO molecules.  
It is suggested that the depletion of carbon and oxygen into dust grains and CO molecules can bring the chemical abundances of the recent ejecta in HuBi\,1 close to those of its [WC] CSPN.  

\end{itemize}

\section*{Acknowledgments}

B.M.M.\ and M.A.G.\ are funded by the Spanish Ministerio de Ciencia, Innovaci\'on y Universidades (MCIU) grant PGC2018-102184-B-I00, co-funded by FEDER funds. 
B.M.M., B.P., M.A.G.\ and S.C.\ acknowledge support from the State Agency for Research of the Spanish MCIU through the ‘Center of Excellence Severo Ochoa’ award to the Instituto de Astrof\'\i sica de Andaluc\'\i a (SEV-2017-0709). 
J.A.T. is funded by UNAM DGAPA PAPIIT project IA100720 and the Marcos Moshinsky Fundation (Mexico). 

This work has made used of observations made with the Gran Telescopio Canarias (GTC), installed at the Spanish Observatorio del Roque de los Muchachos of the Instituto de Astrof\'\i sica de Canarias, in the island of La Palma. 
This work made use of {\sc iraf}, distributed by the National Optical Astronomy Observatory, which is operated by the Association of Universities for Research in Astronomy under cooperative agreement with the National Science Foundation. 
This work has made extensive use of NASA’s Astrophysics Data System.

\section*{Data availability}

The data underlying this article will be shared on reasonable request to the corresponding author.


\bsp	
\label{lastpage}

\appendix

\section{Selection and correction of emission lines in OSIRIS data}
\label{sec:OSIRIS_data}

The emission of the inner shell of HuBi\,1 is contaminated by that of the surrounding outer shell.  
This is a typical situation in multiple shell PNe where the contribution of the emission from the outer shell to the inner shell is normally ignored because the emission from the inner shell is much brighter than that of the outer shell. 
This is not the case for the emission from the inner shell of HuBi\,1, for which the emission from some lines (e.g., the H Balmer lines) is much fainter than that from the outer shell.  
Therefore the contribution of the emission from the outer shell to the inner shell cannot be ignored and the measurement of the emission line fluxes for the inner and outer shells of HuBi\,1 require a tailored approach.

The OSIRIS data provide information on the spatial location of the emission and cannot be used to disentangle the fraction of the emission at the location of the inner shell that corresponds to the inner and outer shells.  
Fortunately, the kinematic information provided by the MEGARA data can be used to separate the emission of the inner shell from that of the outer shell given their different expansion velocities \citep{Rechy}.   
Accordingly, we have defined four different types of emission lines: purely internal, purely external, external with some inner emission, and internal with some external emission. 
The classification for each emission line detected in HuBi\,1 is presented in Table~\ref{tbl:shift_cla}. 
\begin{itemize}
\item 
Purely internal (PI) or external (PE) emission lines. \\ 
Lines whose emission arises mostly from the inner or outer shell, such as the [O~{\sc iii}] or He~{\sc ii} and the He~{\sc i} lines, respectively.  
The fluxes are measured directly from the spectra extracted from the corresponding apertures overlaid on Figure~\ref{fig:fov_int}, i.e, the top and bottom spectra in Figure~\ref{fig:spec_osi}, but for the emission lines with available GTC MEGARA images (e.g., the [O~{\sc iii}], He~{\sc ii} and He~{\sc i} emission lines), whose total integrated fluxes are computed from these images.
\item 
Lines mostly from the external shell with internal emission (EwI). \\
The emission of the outer shell is dominated by H Balmer and He~{\sc i} recombination lines, implying strong contamination to the emission of the inner shell in these lines.  
The different expansion velocity of the inner and outer shells and the tomographic capability of the GTC MEGARA data have allowed us to disentangle the emission in the \ha and \hb lines as described in Sections~\ref{sec:isolation} and \ref{sec:isolation-gaussian}. 
The flux of these emission lines from both shell is thus measured directly on the GTC MEGARA images of each shell. 
In the case of He I lines, with only GTC OSIRIS data, only 5875 He I line had enough S/N to obtain the emission of the inner shell by using lines profile. As well as in H$\beta$ and H$\alpha$ lines profiles has slight bumps at the location of inner shell corresponding with the emission of this shell, He I should has this bumps and we use H$\beta$ to extract the inner emission of the He I line.
First, the dereddened H$\beta$ profile of inner shell (see, section \ref{sec:isolation}) was subtracted from the total profile in order to obtain non-contaminated emission of the outer shell. Then the peaks of total emission He I profile and H$\beta$ pure outer emission profile were equated to estimated the pure emission of He I line. Finally we subtract the H$\beta$ profile from He I to obtain the emission purely internal of the later. Although this method is subject to a worse S/N, it provide an estimation of He I 5875 line. A similar method was also applied to other He I lines however the S/N was too large to obtain emission of the inner shell.
Therefore these lines were assumed to be PE.

\item 
Lines mostly from the internal shell with external emission (IwE). \\
A number of important emission lines are notably bright in the inner shell, but the outer shell still presents significant emission.  
In the case of the emission lines of [N~{\sc i}] and [O~{\sc ii}], with only GTC OSIRIS data, the surface brightness of the outer shell (top panel Figure \ref{fig:spec_osi}) in these lines was measured and an average value adopted on the assumption that it is an homogeneous shell. 
Then, that emission was subtracted from the emission of the inner shell (bottom panel Figure~\ref{fig:spec_osi}) considering  an equivalent aperture for the inner shell to obtain its net flux. 
The flux for outer shell was computed scaling its average surface brightness to the yellow and red apertures in Figure~\ref{fig:fov_int} to consider the flux of the outer shell projected onto the inner shell. 
In the case of the emission lines of [N~{\sc ii}] and [S~{\sc ii}], which have GTC MEGARA data, a similar procedure to that used for the \ha and \hb lines could have been applied. 
However, the emissions from the inner and outer shells of [N~{\sc ii}] and [S~{\sc ii}] could not be perfectly separated in the velocity space, as was done for \ha and \hbs, because the emission from the inner shell is present in all velocity channels, not only in high-velocity components, whereas that of the outer shell has emission from an intermediate region occupying velocity channels with relatively high expansion velocities \citep{Jesus}. 
In this case, we first determined the emission of these lines from the inner shell that could be isolated in high-velocity components as for H$\alpha$ and H$\beta$. 
Then at lower velocities, where intermediate/outer and inner shells are present, a mask of 2.5 arcsec in radius was applied and the flux measured on this aperture.  
The flux of the inner shell was the sum of both terms, whereas for the outer shell, the flux adopted was that of the emission beyond 2.5 arcsec.


\end{itemize}

We note that the \hb emission line 
is used to compute the different line ratios listed in Table~\ref{tbl:t_f_OSI}.  
For the inner shell, these are computed using the total \hb and line fluxes derived from the GTC MEGARA image if there is available GTC MEGARA image of the inner shell for that line.  
If there were no available GTC MEGARA image of that line, the \hb flux is computed from a pseudo-slit onto the GTC MEGARA image at the location of the GTC OSIRIS slit (see next Appendix).


\begin{table}
\begin{tabular*}{0.80\columnwidth}{lrc}
\hline
\multicolumn{1}{l}{Line} & 
\multicolumn{1}{c}{DCR shift} & 
\multicolumn{1}{c}{Line Classification} \\
\multicolumn{1}{c}{} & 
\multicolumn{1}{c}{(arcsec)} & 
\multicolumn{1}{c}{} \\
\hline
[O~{\sc ii}] 3727                    & $+$1.22~~~ & IwE     \\
{[}Ne~{\sc iii}{]} 3869              & $+$1.08~~~ & PI      \\
H$\zeta$+He~\sc{i} 3889              & $+$1.06~~~ & PE      \\
H$\epsilon$+[Ne~\sc{iii}] 3969       & $+$0.99~~~ & $\dots$ \\
He~{\sc i} 4026                      & $\dots$~~~ & PE      \\
{[}S~{\sc ii}{]} 4069                & $+$0.91~~~ & PI      \\
H$\delta$ 4101                       & $+$0.88~~~ & PE      \\
H$\gamma$ 4340                       & $+$0.71~~~ & PE      \\
{[}O~\sc{iii}{]} 4363                & $+$0.70~~~ & PI      \\
He~\sc{i} 4471                       & $+$0.63~~~ & PE      \\
C~{\sc i}:O~{\sc ii}:N~{\sc ii}:4562+4570 & $+$0.58~~~ & PI      \\ 
He~\sc{ii} 4686                      & $+$0.51~~~ & PI      \\
H$\beta$ 4861                        & $+$0.43~~~ & EwI     \\
{[}O~\sc{iii}{]} 4959                & $+$0.38~~~ & PI      \\
{[}O~\sc{iii}{]} 5007                & $+$0.36~~~ & PI      \\
He~\sc{i} 5016                       & $+$0.36~~~ & PE      \\
{[}N~\sc{i}] 5199                    & $+$0.29~~~ & IwE     \\
{[}N~\sc{ii}] 5755                   & $+$0.11~~~ & PI   \\
He~\sc{i} 5876                       & $+$0.08~~~ & EwI   \\
{[}O~\sc{i}] 6300                    & $+$0.01~~~ & PI   \\
{[}S~\sc{iii}] 6312                  & $-$0.01~~~ & PI   \\
{[}O~\sc{i}] 6363                    & $-$0.02~~~ & PI   \\
{[}N~\sc{ii}] 6548                   & $-$0.05~~~ & IwE  \\
H$\alpha$ 6563                       & $-$0.05~~~ & EwI  \\
{[}N~\sc{ii}] 6584                   & $-$0.06~~~ & IwE  \\ 
He~\sc{i} 6678                       & $-$0.07~~~ & PE   \\
{[}S~\sc{ii}] 6717                   & $-$0.08~~~ & IwE  \\
{[}S~\sc{ii}] 6731                   & $-$0.08~~~ & IwE  \\
He~\sc{i} 7065                       & $-$0.13~~~ & PE   \\
{[}Ar~\sc{iii}] 7135                 & $-$0.14~~~ & PI   \\
C~\sc{ii} 7236                       & $-$0.16~~~ & PI   \\
C~\sc{i}, O~\sc{ii}? 7289            & $-$0.16~~~ & PI   \\
{[}O~\sc{ii}] 7320                   & $-$0.17~~~ & PI   \\
{[}O~\sc{ii}] 7330                   & $-$0.17~~~ & PI   \\
N~\sc{i}:O~\sc{ii} 7379            & $-$0.18~~~ & PI   \\
\hline
\end{tabular*}
\caption{
Atmospheric differential chromatic refraction (DCR) shift for the emission lines detected in the GTC OSIRIS spectrum of HuBi\,1 and line classification according to the fraction of emission from the inner and outer shells. 
}
\label{tbl:shift_cla}
\end{table}

\section{Correcting the Atmospheric Differential Chromatic Refraction in the OSIRIS Data} 
\label{sec:chromatic}

The OSIRIS observations of HuBi\,1 were obtained at the time around its culmination, but still at large airmass values at the ORM, in the range from 1.49 to 1.41, when notable differential chromatic refraction (DCR) effects can be expected. 
DCR effects are usually minimised selecting the position angle (PA) of the slit along the parallactic angle, which is close to the North-South direction near culmination. 
The PA of the slit during the observations was however set almost orthogonally, at PA 90$^\circ$ along the East-West direction, to avoid background stars and to register nebular areas of the highest surface brightness, thus optimising the scientific return of the observations.  
As a result, DCR effects cannot be ignored and need to be accounted for in the analysis of this data set.

According to \citet{Filippenko1982}, the positional shift along the parallactic angle caused by the DCR can be expressed in terms of $\lambda$ and zenith distance $z$ (or airmass $m$), which is a function of the observing time $t$.  
This shift is computed for a wavelength reference at 5000 \AA\ adopting typical conditions of the ORM. 
We have computed the spatial shifts for the wavelength of each emission line of interest according to Filippenko's expressions, but referred them to the central wavelength of the Sloan \emph{r'} filter at 6204 \AA\ that was used for the acquisition of HuBi\,1. 

\begin{equation}
\mathrm{DCR}(\lambda) = R(\lambda)- R(\lambda_0) \approx 206265[n(\lambda) - n(\lambda_0)] \tan(z) \end{equation}
\begin{equation}
n(\lambda)_{15,760}-1 = \left[ 64.328 + \frac{29498.1}{146-(1/\lambda)^2}+ \frac{255.4}{41-(1/\lambda)^2} \right]
\end{equation}
\begin{equation}
n(\lambda)_{T,P} -1 = (n(\lambda) 15,760 -1) \\
\times \frac{P[1+(1.049- 0.0157T)  10^{-6} P]}{720.883(1+0.003661T)}
\end{equation}

The positional shift between emission lines at the blue and red extremes of the spectral range (for instance, the [O~{\sc ii}] $\lambda$3727 and [O~{\sc ii}] $\lambda\lambda$7320,7330 emission lines) was found to be larger than the slit width.  

\begin{figure}
\centering
\includegraphics[clip,width=1\columnwidth]{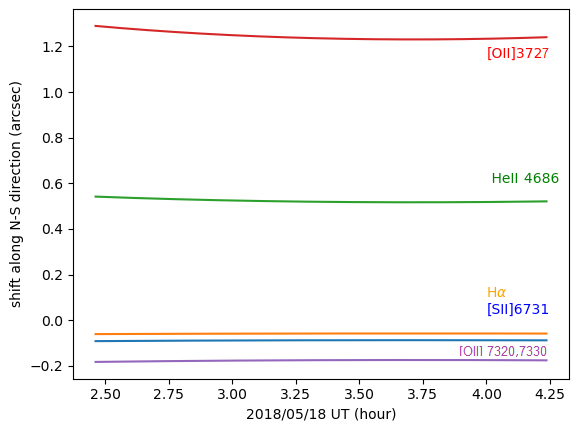}
\caption{Variations of the shifts with respect to [N~{\sc ii}] for different emission lines during the observations. DCR becomes noticeable in the blue zone of spectrum rising up 1.2 arcsec for [O~{\sc ii}]$\lambda$3727. Due to the negligible variation, an average value of the shifted for each line can be adopted for the whole observation.}\label{fig:dcr}
\end{figure}

Since the parallactic angle is a function of time $pa(t)$ that varies quite rapidly near culmination, the shift of the slit along the orthogonal direction caused by DCR effects $s(t)$ needs to be corrected accordingly: 
\begin{equation}
s(t) = DCR(\lambda,t) \times \cos pa(t)
\end{equation}
The variations of the shifts with respect to [N~{\sc ii}] for different emission lines of interest such as [O~{\sc ii}] $\lambda$3727, He~{\sc ii} $\lambda$4686, H$\beta$, [O~{\sc iii}] $\lambda$5007, [S~{\sc ii}] $\lambda\lambda$6716,6731, and [O~{\sc ii}] $\lambda\lambda$7320,7330 are shown in Figure \ref{fig:dcr}.  
This figure reveals that the variations of the shifts during the observations are not dramatic for the wavelength of each emission lines of interest.

\begin{figure}
\centering
\includegraphics[clip,width=1\columnwidth]{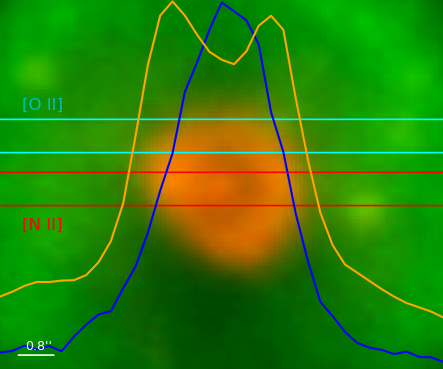}
\caption{NOT ALFOSC [N~{\sc ii}] $\lambda$6584 (red) and H$\alpha \lambda$6563 (green) colour composite picture of HuBi 1. Relative positions of OSIRIS slit for [O~{\sc ii}] (cyan) and [N~{\sc ii}] (red) are superimposed and also its corresponding lines profiles (blue) and (orange), respectively. The relative position of [O~{\sc ii}] line falls just on the edge of the inner region, providing and single-peak profile, and not two as would be expected looking at figure \ref{fig:total_flux}, showing the DCR suffered by OSIRIS data set.}\label{fig:shift_dcr}
\end{figure}

We can thus define slit positions for each emission line using median values of their shifts in Figure \ref{fig:dcr} to characterise the spatial regions of HuBi\,1 registered by these emission lines. 
A number of these equivalent slit positions are overlaid on an image of HuBi\,1 in the [N~{\sc ii}] $\lambda$6584 emission line in Figure~\ref{fig:shift_dcr}.  
The figure immediately reveals that different emission lines register different nebular regions.  
This is particularly significant for the small-sized inner shell of HuBi\,1, implying that the comparison of spatial profiles and fluxes for different emission lines of this shell extracted from the OSIRIS 2D spectra has to consider these shifts \citep[unlike the analysis presented by][]{Pena2021}.

In order to correct from these effects the intensity line ratio of a particular emission line to H$\beta$, we have adopted different strategies.  
For the inner shell, the intensity line ratio to H$\beta$ of emission lines covered in the spectral range of the MEGARA data is derived from the whole emission of the inner shell, whereas that of emission lines only covered in the OSIRIS data are compared to the H$\beta$ flux derived from the corresponding pseudo-slit of this line in the MEGARA data.  
For the outer shell, the MEGARA data do not cover the whole outer shell, thus the intensity line ratio to H$\beta$ has been solely derived using the latter approach, i.e., the flux of the emission line is compared to the H$\beta$ flux derived from the corresponding pseudo-slit of this line in the MEGARA data.  
This procedure warranties that the intensity line ratio to H$\beta$ of a particular emission line is derived using line fluxes measured in the same nebular apertures for both emission lines.

\section{{\sc mappings v} shock models of the inner shell of HuBi\,1}
\label{sec:app_mappings}

{\sc mappings v} \citep[see][]{Sutherland_2017} is able to produce line and continuum emissivity from shocked plasmas from  non-equilibrium ionisation objects, for example, radiative shocks in supernova remnants or shock structures surrounding  Herbig-Haro objects. 
{\sc mappings v} solves the coupled ionisation and cooling equations in a time-dependent scheme. 
The first version of {\sc mappings} \citep[see][]{Binette1985} was developed as a result of a series of works presented by \citet{Dopita1976,Dopita_1977,Dopita1978}. 
The code has been greatly improved since then and it now includes a detailed treatment of temperature-dependent collision strengths, new cooling functions, computations for optically thin plasmas and up to 80,000 cooling and recombination lines \citep[see the historical improvement of the code presented in][]{Sutherland_2017}.

Here we present in Tables~\ref{tab:c1} to \ref{tab:c4}
the predicted line intensities of some of the models computed with {\sc mappings v} following the methodology explained in Sec.~\ref{sec:abundancias} to estimate the chemical abundances of the inner shell of HuBi\,1. 
All models are computed with a spectral resolution $R \approx$ 3000, i.e.\ the line intensity ratios presented in Tables~\ref{tab:c1} to \ref{tab:c4} include the contribution of all emission lines computed in the model around the nominal wavelength that would be unresolved at this spectral resolution.  
We note that the expected contribution of contaminant emission lines is mostly negligible.  

Tables~\ref{tab:c1} and \ref{tab:c2} present emission line ratios referred to $I$(H$\beta$)=100 as they constrain the chemical abundances of helium, oxygen, and nitrogen with respect to hydrogen. 
The first five models in Table~\ref{tab:c1} were computed at fixed values of He/H and N/H to explore the variations of the emission line ratios with the O abundances.  
Then the O/H abundances ratio was fixed at its best value and the N/H abundances ratio varied in the next five columns. 
This table shows that high oxygen and nitrogen abundances are necessary to reproduce the emission lines of [O~{\sc iii}] and [N~{\sc ii}]. 
Once the best-fit values of the O/H and N/H abundances ratios were determined, the He/H abundances ratio was varied in Table~\ref{tab:c2}. 
Although the He~{\sc i} $\lambda$5876 emission line is not well reproduced by the adopted value of He/H, we note that different values of He/H lead to notable differences in the estimate of most emission lines (being clearly overestimated for 12+log(He/H) > 13.0).

Tables~\ref{tab:c3} and \ref{tab:c4} explore the variations of the intensity line ratios of He, N, O, Ne, S, and Ar induced by changes of the physical conditions of the shock, namely the shock velocity $v_{\rm s}$ and the pre-shock density $n_{\rm pre}$. 
We therefore preferred to refer the emission line ratios to $I$({[O~{\sc iii}]} 5007)=100 to assess the relative changes in the line ratios from ions of those elements.  
It should be noted in Table~\ref{tab:c3} that the [O~{\sc ii}] to [O~{\sc iii}] line ratio peak at $v_{\rm s}$ of 100 km~s$^{-1}$, then drops gradually at higher velocities.  
Indeed, the notable increase of the [O~{\sc iii}] line intensities at velocities 140 km~s$^{-1}$ and above produces a sharp decline in most line ratios, particularly in the H~{\sc i} , He~{\sc i}, and He~{\sc ii} recombination lines. 
Meanwhile the effects of the variations of the pre-shock density $n_{\rm pre}$ explored in Table~\ref{tab:c4} reveal notable changes in the intensity ratios of lines from low ionization species, such as He~{\sc i}, [O~{\sc i}], [O~{\sc ii}], [N~{\sc i}], [N~{\sc ii}], and [S~{\sc ii}], with respect to [O~{\sc iii}], whereas other higher ionization species show less dramatic variations.



\begin{table*}
\caption{
{\sc mappings v} models assessing the variation of the oxygen and nitrogen abundances for the inner shell of HuBi\,1.  
The values of $v_\mathrm{s} = 80$ km~s$^{-1}$, $n_\mathrm{pre}=1$~cm$^{-3}$, $B=1\,\mu$G, 12+log$_{10}$(He/H)=13.0, and H/H$^+$=He/He$^+$=1 have been fixed in all models. 
O/H stands for 12+log$_{10}$(O/H), N/H for 12+log$_{10}$(N/H), and RO3 denotes the ratio [O\,{\sc iii}] $\lambda$4363/[O\,{\sc iii}] $\lambda$5007.  
}
\setlength{\tabcolsep}{0.8\tabcolsep}
\label{tab:c1}
\begin{tabular}{lr|rrrrr|rrrrr}
\hline
Line        & Obs.         & 
\multicolumn{5}{c}{\underline{\hspace*{2.7cm} N/H=7.9 \hspace*{2.7cm}}} & 
\multicolumn{5}{c}{\underline{\hspace*{2.7cm} O/H=9.8 \hspace*{2.7cm}}} \\
& & 
\multicolumn{1}{c}{O/H=8.0}  & 
\multicolumn{1}{c}{O/H=9.0}  & 
\multicolumn{1}{c}{O/H=9.8}  & 
\multicolumn{1}{c}{O/H=10.0}  & 
\multicolumn{1}{c|}{O/H=10.2}  & 
\multicolumn{1}{c}{N/H=7.9}  & 
\multicolumn{1}{c}{N/H=8.9} & 
\multicolumn{1}{c}{N/H=9.9} & 
\multicolumn{1}{c}{N/H=10.9} & 
\multicolumn{1}{c}{N/H=11.9} \vspace*{0.05cm} \\ 
\hline 
{[O~{\sc ii}]} 3727           &       8360 &           166.6 &           723.2 &          1049.7 &           978.6 &           985.3 &          1049.7 &           968.8 &           571.2 &            66.5 &           398.1\\   
{[Ne~{\sc iii}]} 3869        &        622 &             1.7 &            15.6 &            54.6 &            97.0 &            88.4 &            54.6 &            74.9 &            54.9 &            13.0 &            63.8\\
{[S~{\sc ii}]} 4069         &        111 &             5.2 &            20.4 &            21.1 &            24.0 &            21.7 &            21.1 &            23.7 &            14.2 &             2.1 &            37.4\\ 
{[O~{\sc iii}]} 4363          &         30.6 &             0.7 &             6.1 &            22.8 &            44.2 &            43.7 &            22.8 &            31.1 &            25.0 &             6.3 &            28.3\\ 
He~{\sc ii} 4685              &        134 &            43.8 &            48.2 &            49.2 &            75.3 &            60.6 &            49.2 &            70.6 &            66.8 &            49.7 &            69.0\\ 
{[O~{\sc iii}]} 4959   &       101 &             3.2 &            26.8 &           109.9 &           229.0 &           252.6 &           109.9 &           149.1 &           129.2 &            45.1 &           143.3\\
{[O~{\sc iii}]} 5007     &     324 &             9.2 &            77.7 &           317.8 &           662.1 &           730.1 &           317.8 &           431.0 &           373.4 &           130.4 &           414.5\\  
{[N~{\sc i}]} 5199           &         75.6 &            58.4 &            18.7 &             2.1 &             1.1 &             0.7 &             2.1 &            22.1 &           119.3 &           266.9 &           111.3\\ 
{[N~{\sc ii}]} 5755           &       60.2 &             9.2 &             4.4 &             1.1 &             0.7 &             0.5 &             1.1 &            10.4 &            66.3 &            85.7 &            88.4\\ 
He~{\sc i} 5876                   & <           20.5 &           286.0 &           361.8 &           468.9 &           399.5 &           322.6 &           468.9 &           422.7 &           336.7 &           165.6 &           388.1\\  
{[O~{\sc i}]} 6300      &      95.5 &            27.2 &            88.9 &            51.1 &            37.7 &            33.9 &            51.1 &            56.6 &            30.4 &             5.3 &            76.5\\   
{[S~{\sc iii}]} 6312          &        4.5 &             0.6 &             3.4 &             7.1 &             9.2 &             9.4 &             7.1 &             7.6 &             5.3 &             1.1 &             5.7\\ 
{[O~{\sc i}]} 6363             &         28.3 &             8.7 &            28.4 &            16.3 &            12.1 &            10.8 &            16.3 &            18.1 &             9.7 &             1.7 &            24.5\\
{[N~{\sc ii}]} 6548           &        330 &           126.6 &            45.2 &             8.1 &             5.6 &             3.4 &             8.1 &            86.9 &           552.2 &           871.3 &           754.3\\ 
H$\alpha$             &           274.0 &           323.5 &           325.2 &           397.3 &           410.0 &           436.3 &           397.3 &           392.4 &           407.4 &           472.3 &           361.9\\ 
{[N~{\sc ii}]} 6584           &       1390 &           372.6 &           132.9 &            24.0 &            16.4 &            10.0 &            24.0 &           255.6 &          1624.6 &          2563.6 &          2219.4\\ 
{[S~{\sc ii}]} 6716           &        135 &            62.3 &           181.2 &           160.4 &           144.5 &           133.7 &           160.4 &           172.2 &           104.0 &            18.1 &            75.7\\
{[S~{\sc ii}]} 6731           &        112 &            50.3 &           168.1 &           163.7 &           159.5 &           128.7 &           163.7 &           190.4 &           115.0 &            17.4 &           117.4\\ 
{[Ar~{\sc iii}]} 7135        &         11.1 &             0.5 &             3.7 &             8.3 &            13.8 &            12.9 &             8.3 &            11.0 &             7.8 &             1.7 &             9.2\\
{[O~{\sc ii}]} 7319           &         31.6 &             7.4 &            42.9 &            70.7 &            73.9 &            62.5 &            70.7 &            73.8 &            43.1 &             3.7 &            88.8\\  
{[O~{\sc ii}]} 7329           &         26.2 &             6.0 &            34.7 &            57.1 &            59.7 &            50.5 &            57.1 &            59.6 &            34.8 &             3.0 &            72.0\\ \hline 
RO3 &          0.094 &          0.076 &          0.078 &          0.072 &          0.067 &          0.060 &          0.072 &          0.072 &          0.067 &          0.048 &          0.068\\ 
\hline
\end{tabular}
\end{table*}


\newpage

\begin{table*}
\caption{{\sc mappings v} models assessing the variation of the helium abundances in the inner shell of HuBi\,1. 
The values of $v_\mathrm{s} = 80$ km~s$^{-1}$, $n_\mathrm{pre} = 1$ cm$^{-3}$, $B = 1\,\mu$G, 12+log$_{10}$(O/H)=9.8, 12+log$_{10}$(N/H)=9.9, and H/H$^+$=He/He$^+$=1 have been fixed in all models. 
He/H stands for 12+log$_{10}$(He/H) and RO3 denotes the ratio [O\,{\sc iii}] $\lambda$4363/[O\,{\sc iii}] $\lambda$5007.
}
\label{tab:c2}
\begin{tabular}{lrrrrrr}
\hline
& & \multicolumn{5}{c}{\underline{\hspace*{2.4cm} He/H \hspace*{2.4cm}}} \\
Line                   & Obs.         &         12.2 &         12.9 &         13.0 &         13.1 &         13.2 \\
\hline 
{[O~{\sc ii}]} 3727           &       8360.0 &        185.7 &        493.7 &        571.2 &        652.6 &        721.9\\
{[Ne~{\sc iii}]} 3869        &        622.0 &         38.7 &         58.8 &         54.9 &         50.2 &         49.9\\
{[S~{\sc ii}]} 4069         &        111 &          6.4 &         12.0 &         14.2 &         16.9 &         21.1\\ 
{[O~{\sc iii}]} 4363          &         30.6 &         20.6 &         27.9 &         25.0 &         21.8 &         20.5\\
He~{\sc ii} 4685              &        134.0 &         19.1 &         66.0 &         66.8 &         66.7 &         73.0\\
{[O~{\sc iii}]} 4959          & 101.0 &        129.4 &        149.8 &        129.2 &        108.6 &         98.5\\
{[O~{\sc iii}]} 5007     &     324.0 &        374.2 &        433.0 &        373.4 &        314.1 &        284.9\\ 
{[N~{\sc i}]} 5199           &         75.6 &        221.4 &         90.6 &        119.3 &        152.1 &        190.1\\  
{[N~{\sc ii}]} 5755           &       60.2 &         17.2 &         58.8 &         66.3 &         74.3 &         83.3\\  
He~{\sc i} 5876                      &   <        20.5 &         60.8 &        299.7 &        336.7 &        395.9 &        474.7\\
{[O~{\sc i}]} 6300            &         95.5 &         44.6 &         20.4 &         30.4 &         42.2 &         57.9\\ 
{[S~{\sc iii}]} 6312   &         4.4 &          3.0 &          5.4 &          5.3 &          5.4 &          5.6\\
{[O~{\sc i}]} 6363            &         28.3 &         14.2 &          6.5 &          9.7 &         13.5 &         18.5\\ 
{[N~{\sc ii}]} 6548    &        330 &        285.6 &        485.5 &        552.2 &        622.0 &        718.5\\ 
H$\alpha$             &        274.0 &        487.2 &        406.6 &        407.4 &        408.0 &        404.5\\ 
{[N~{\sc ii}]} 6584    &       1390 &        840.3 &       1428.5 &       1624.6 &       1830.1 &       2114.2\\ 
{[S~{\sc ii}]} 6716    &        135 &         101.7 &         87.0 &        104.0 &        119.7 &        137.3\\  
{[S~{\sc ii}]} 6731    &        112 &         72.8 &         92.8 &        115.0 &        136.3 &        164.4\\ 
{[Ar~{\sc iii}]} 7135    &        11.1 &          5.3 &          8.2 &          7.8 &          7.3 &          7.4\\ 
{[O~{\sc ii}]} 7319    &         31.6 &          6.3 &         35.0 &         43.1 &         51.7 &         62.5\\
{[O~{\sc ii}]} 7329    &         26.2 &          5.1 &         28.3 &         34.8 &         41.8 &         50.4\\
\hline 
RO3 &          0.094 &          0.055 &          0.065 &          0.067 &          0.070 &          0.072\\  \hline
\end{tabular}
\end{table*}

\begin{table*}
\caption{
{\sc mappings v} models assessing the variation of the shock velocity $v_{\rm s}$.  
The values of $n_\mathrm{pre} =1$~cm$^{-3}$, $B = 1\,\mu$G, 12+log$_{10}$(O/H)=9.8, 12+log$_{10}$(N/H)=9.9, 12+log$_{10}$(He/H)=13.0, and H/H$^+$=He/He$^+$=1 have been fixed in all models. 
}
\label{tab:c3}
\begin{tabular}{lrrrrrrrr}
\hline
& & \multicolumn{7}{c}{\underline{\hspace*{2.8cm} $v_\mathrm{s}$ (km~s$^{-1}$) \hspace*{2.8cm}}} \\
Line                 & \multicolumn{1}{c}{Obs.}           & 80 & 100 & 120 & 140 & 160 & 180 & 200 \\
\hline 
{[O~{\sc ii}]} 3727           &       2580.2 &           153.0 &           495.2 &           219.4 &            50.5 &            22.6 &            13.6 &             9.0\\  
{[Ne~{\sc iii}]} 3869        &        192.0 &            14.7 &            10.1 &            13.5 &             6.5 &             7.0 &             6.8 &             6.5\\ 
{[S~{\sc ii}]} 4069         &        34.3 &             3.8 &             6.0 &             3.6 &             0.8 &             0.4 &             0.4 &             0.3\\ 
{[O~{\sc iii}]} 4363          &          9.4 &             6.7 &             6.5 &             7.8 &             2.6 &             3.5 &             4.0 &             4.1\\ 
He~{\sc ii} 4685              &        41.4 &            17.9 &            13.2 &            17.7 &             1.2 &             1.5 &             2.5 &             3.4\\ 
H$\beta $          &            30.9 &            26.8 &            54.8 &            43.5 &             0.3 &             0.4 &             1.0 &             0.6\\ 
{[O~{\sc iii}]} 4959          &  31.2 &            34.6 &            34.6 &            34.6 &            34.6 &            34.6 &            34.6 &            34.6\\ 
{[O~{\sc iii}]} 5007     &      100.0 &           100.0 &           100.0 &           100.0 &           100.0 &           100.0 &           100.0 &           100.0\\ 
{[N~{\sc i}]} 5199           &         23.3 &            31.9 &            64.3 &            24.9 &             3.2 &             2.2 &             2.3 &             1.5\\   
{[N~{\sc ii}]} 5755           &       18.6 &            17.7 &            50.8 &            23.9 &             5.4 &             2.8 &             1.8 &             1.3\\   
He~{\sc i} 5876                      &   <            6.3 &            90.2 &           131.7 &            63.9 &             2.4 &             1.7 &             2.4 &             1.4\\ 
{[O~{\sc i}]} 6300            &         29.5 &             8.1 &            13.8 &             7.8 &             0.7 &             0.5 &             0.8 &             0.4\\  
{[S~{\sc iii}]} 6312   &         1.4 &             1.4 &             3.2 &             1.8 &             0.7 &             0.8 &             0.8 &             0.8\\ 
{[O~{\sc i}]} 6363            &         8.7 &             2.6 &             4.4 &             2.5 &             0.2 &             0.2 &             0.2 &             0.1\\ 
{[N~{\sc ii}]} 6548    &         101.9 &           147.9 &           296.6 &           183.8 &            36.0 &            15.4 &            13.6 &             6.9\\ 
H$\alpha$         & 84.5 &           109.1 &           245.5 &           175.6 &             1.0 &             1.7 &             3.8 &             2.2\\ 
{[N~{\sc ii}]} 6584    &      429.0 &           435.0 &           872.7 &           540.9 &           106.0 &            45.4 &            40.1 &            20.2\\ 
{[S~{\sc ii}]} 6716    &        41.7 &            27.9 &            48.9 &            26.4 &             6.7 &             3.2 &             2.8 &             1.7\\
{[S~{\sc ii}]} 6731    &        34.6 &            30.8 &            41.9 &            29.5 &             4.8 &             2.3 &             2.7 &             1.3\\ 
{[Ar~{\sc iii}]} 7135    &        3.4 &             2.1 &             2.0 &             1.7 &             0.9 &             0.6 &             0.5 &             0.4\\ 
{[O~{\sc ii}]} 7319    &         9.8 &            11.5 &            28.6 &            16.0 &             2.2 &             1.1 &             0.8 &             0.5\\ 
{[O~{\sc ii}]} 7329    &          8.1 &            9.3 &            23.2 &            13.0 &             1.8 &             0.9 &             0.6 &             0.4\\ \hline 
\end{tabular}
\end{table*}

\begin{table*}
\caption{
{\sc mappings v} models assessing the variation of the pre-shock density $n_{\rm pre}$.  
The values of $v_\mathrm{s}=80$~km~s$^{-1}$, $B =1\,\mu$G, 12+$\log$(O/H)=9.8, 12+$\log$(N/H)=9.9, 12+$\log$(He/H)=13.0, and H/H$^+$=He/He$^+$=1 have been fixed in all models. 
}
\label{tab:c4}
\begin{tabular}{lrrrrr}
\hline
& & \multicolumn{4}{c}{\underline{\hspace*{1.2cm} $n_\mathrm{pre}$ (cm$^{-3}$) \hspace*{1.2cm}}} \\
Line                 & Obs.         &         0.1 &         1.0 &         10 &         100 \\
\hline 
{[O~{\sc ii}]} 3727           &       2580.2 &          1104.2 &           153.0 &           693.9 &            98.7\\ 
{[Ne~{\sc iii}]} 3869        &        192.0 &            11.5 &            14.7 &            12.6 &            17.0\\  
{[S~{\sc ii}]} 4069         &        34.3 &             9.8 &             3.8 &            13.6 &            21.7\\
{[O~{\sc iii}]} 4363          &         9.4 &             6.3 &             6.7 &             6.3 &             7.7\\
He~{\sc ii} 4685              &        41.4 &            13.2 &            17.9 &            11.6 &            15.9\\
H$\beta$          &            30.9 &            16.6 &            26.8 &            93.3 &            25.8\\ 
{[O~{\sc iii}]} 4959          &  31.2 &            34.6 &            34.6 &            34.6 &            34.5\\
{[O~{\sc iii}]} 5007     &      100.0 &           100.0 &           100.0 &           100.0 &           100.0\\
{[N~{\sc i}]} 5199           &         23.3 &            45.4 &            31.9 &           124.1 &            20.6\\   
{[N~{\sc ii}]} 5755           &       18.6 &           100.2 &            17.7 &            72.7 &            40.7\\  
He~{\sc i} 5876                      &   <            6.3 &            47.0 &            90.2 &           444.8 &           107.6\\ 
{[O~{\sc i}]} 6300            &         29.5 &            10.0 &             8.1 &            32.0 &            44.6\\ 
{[S~{\sc iii}]} 6312   &         1.4 &             6.1 &             1.4 &             3.9 &             1.8\\ 
{[O~{\sc i}]} 6363            &         8.7 &             3.2 &             2.6 &            10.2 &            14.3\\
{[N~{\sc ii}]} 6548    &        101.9 &           611.1 &           147.9 &           461.8 &           199.4\\
H$\alpha $         &            84.5 &           68.7 &           109.1 &           371.2 &           84.8\\ 
{[N~{\sc ii}]} 6584    &       429.0 &          1797.9 &           435.0 &          1358.7 &           586.6\\ 
{[S~{\sc ii}]} 6716    &        41.7 &            87.4 &            27.9 &            79.7 &            13.7\\ 
{[S~{\sc ii}]} 6731    &        34.6 &            62.1 &            30.8 &            68.2 &            18.0\\ 
{[Ar~{\sc iii}]} 7135    &        3.4 &             2.2 &             2.1 &             2.4 &             2.6\\ 
{[O~{\sc ii}]} 7319    &         9.8 &            54.8 &            11.5 &            46.2 &            37.1\\
{[O~{\sc ii}]} 7329    &          8.1 &            44.4 &            9.3 &            37.4 &            30.1\\ \hline
\end{tabular}
\end{table*}


\end{document}